\documentclass[superscriptaddress,longbibliography, reprint,amsmath,amssymb,aps,prl,floatfix]{revtex4-2}
\usepackage[margin=1.5cm]{geometry}

\usepackage{xcolor} 
\usepackage{graphicx}
\usepackage{dcolumn}
\usepackage{upgreek}
\usepackage{bm}
\usepackage{hyperref}
\usepackage{floatrow}
\usepackage{times}
\usepackage[mathlines]{lineno}
\usepackage{physics}
\usepackage{color,soul}
\usepackage{lipsum}  
\usepackage[utf8]{inputenc}
\usepackage{graphicx}
\usepackage{etoolbox}
\usepackage{comment}
\usepackage{multirow}


\graphicspath{{Images/}}

\begin{document}

\title{Demonstration of the Two-Fluxonium Cross-Resonance Gate}

\author{Ebru Dogan}
\affiliation{Department of Physics, University of Massachusetts-Amherst, Amherst, MA, USA}
\author{Dario Rosenstock}
\affiliation{Department of Physics, University of Massachusetts-Amherst, Amherst, MA, USA}
\author{Lo\"ick Le Guevel}
\affiliation{Department of Physics, University of Massachusetts-Amherst, Amherst, MA, USA}
\affiliation{Department of Electrical and Computer Engineering, University of Massachusetts-Amherst, MA, USA}
\author{Haonan Xiong}
\affiliation{Department of Physics, Joint Quantum Institute, and Center for Nanophysics and Advanced Materials, University of Maryland, College Park, MD, USA}
\author{Raymond A. Mencia}
\affiliation{Department of Physics, Joint Quantum Institute, and Center for Nanophysics and Advanced Materials, University of Maryland, College Park, MD, USA}
\author{Aaron Somoroff}
\affiliation{Department of Physics, Joint Quantum Institute, and Center for Nanophysics and Advanced Materials, University of Maryland, College Park, MD, USA}
\author{Konstantin N.~Nesterov}
\affiliation{Department of Physics and Wisconsin Quantum Institute, University of Wisconsin–Madison, Madison, WI, USA}
\author{Maxim G.~Vavilov}
\affiliation{Department of Physics and Wisconsin Quantum Institute, University of Wisconsin–Madison, Madison, WI, USA}
\author{Vladimir E.~Manucharyan}
\affiliation{Department of Physics, Joint Quantum Institute, and Center for Nanophysics and Advanced Materials, University of Maryland, College Park, MD, USA}
\author{Chen Wang}
\email{wangc@umass.edu}
\affiliation{Department of Physics, University of Massachusetts-Amherst, Amherst, MA, USA}

\date{\today}

\begin{abstract}

The superconducting fluxonium qubit has a great potential for high-fidelity quantum gates with its long coherence times and strong anharmonicity at the half-flux quantum sweet-spot. However, current implementations of two-qubit gates compromise fluxonium's coherence properties by 
requiring either a temporary population of the non-computational states or tuning the magnetic flux off the sweet-spot.  Here we realize a fast all-microwave cross-resonance gate between two capacitively-coupled fluxoniums 
with the qubit dynamics well confined to the computational space.  We demonstrate a direct CNOT gate in 70 ns with fidelity up to $\mathcal{F}=0.9949(6)$ despite the limitations of a sub-optimal device coherence and measurement setup. Our results project a possible pathway towards reducing the two-qubit error rate below $10^{-4}$ with present-day technologies. 
\end{abstract}

\maketitle

The spectacular development of superconducting circuits into a leading platform for scaling up quantum computation~\cite{arute_quantum_2019, kjaergaard_superconducting_2020} over the past decade has been almost exclusively riding on the optimization of one type of Josephson qubits: the transmons~\cite{koch_charge-insensitive_2007}.  Among the numerous possibilities on the ``Mendeleev table" of superconducting artificial atoms~\cite{devoret_superconducting_2013}, the transmon has been entrenched as the go-to qubit for reasons that are more practical than fundamental:  It is easy to build with only one or two Josephson junctions, simple to model in the oscillator basis using perturbation theory, and robust to operate with minimal spurious degrees of freedom.  However, it sacrifices anharmonicity, a fundamental quantum resource, for suppression of charge noise, and features a rather restricted parameter space.  Recently, the fluxonium qubit~\cite{manucharyan_fluxonium_2009} emerged as a serious challenger to the monopoly of transmons as the building block of a superconducting quantum processor~\cite{bao_fluxonium_2021, nguyen_scalable_2022}, becoming just the second type of superconducting qubits crossing the 99\% fidelity threshold for two-qubit gates~\cite{ficheux_fast_2021, xiong_arbitrary_2022, bao_fluxonium_2021}. 

Moving forward, fluxonium qubits have the potential to outperform transmons in gate fidelity due to its inherent advantages of having both longer coherence times and higher anharmonicity.  
In particular, the lowest two energy levels of the fluxonium at the half flux quantum, to be used as the computational states $\ket{0}$ and $\ket{1}$, enjoy substantial protection from  dielectric loss due to their low transition frequencies~\cite{nguyen_high-coherence_2019} (typically 100 MHz--1 GHz compared to typical transmons at 4--6 GHz, and can be made even lower~\cite{zhang_universal_2021}), recently reaching a record-setting 1 millisecond in coherence times~\cite{somoroff_millisecond_2021}.  Despite the low qubit frequency, the non-computational transitions to higher excited states are in the range of several GHz, and this strong anharmonicity provides a large bandwidth and on-demand interactions to enable fast gate operations~\cite{nesterov_microwave-activated_2018}.

However, existing implementations of fluxonium two-qubit gates were not yet fully utilizing these core advantages.  One class of CZ or CPhase gates employ the geometric phase imprinted on selected computational states by driving the non-computational $\ket{1}$-$\ket{2}$ transitions~\cite{ficheux_fast_2021, xiong_arbitrary_2022}. These schemes temporarily populate the higher excited states during the gate operation, and therefore are fundamentally limited by their faster (transmon-like) decoherence rates.  
Another prototypical two-qubit gate, the flux-controlled $i\textsc{SWAP}$ gate~\cite{bao_fluxonium_2021}, requires tuning qubit frequencies to activate resonant excitation exchange.  This scheme temporarily brings the fluxonium away from the half-flux ``sweet spot" and is therefore susceptible to first-order flux noise in the same way as similar schemes for transmons.  
A very recent f\textsc{s}im gate using a tunable coupler mitigates but not yet eliminates the need of tuning qubit frequencies away from the half-flux~\cite{moskalenko_high_2022}. 
To fulfil the full potential the fluxonium has to offer, several proposals have been put forward to carry out two-qubit gates within the high-coherence computational subspace at fixed frequencies~\cite{nesterov_proposal_2021, nguyen_scalable_2022, nesterov_controlled-not_2022}, among which the cross-resonance \textsc{CNOT} gate~\cite{nesterov_controlled-not_2022} best leverages the fluxonium anharmonicity.

In the cross-resonance (CR) scheme, a control qubit is strongly driven at the resonance frequency of a (coupled) target qubit, leading to a \textsc{CNOT} operation or its equivalent.  The CR gate have gained substantial popularity in transmon-based systems due to 
its all-microwave implementation and simplistic experimental requirements \cite{rigetti_fully_2010, sheldon_procedure_2016, kandala_demonstration_2021,  heya_cross-cross_2021, corcoles_process_2013, hazra_engineering_2020, patterson_calibration_2019}.  
After a decade of experimental optimization and theoretical modeling~\cite{tripathi_operation_2019, petrescu_accurate_2021, magesan_effective_2020, malekakhlagh_first-principles_2020}, the two-transmon CR gate has reached state-of-the-art fidelity of up to 99.8\%~\cite{wei_quantum_2021}. 
Despite this success, transmons are arguably poorly suited for the CR gate:  In a nutshell, the CR interaction arises when the control qubit acts as a switchable microwave filter which regulates the amplitude and phase of the effective drive field arriving at the target qubit. 
The anharmonicity of the qubit is the crucial resource 
that dictates how strongly this filtering effect depends on the control qubit 
state~\cite{tripathi_operation_2019}.  With limited anharmonicity, the optimized transmon CR gate has to navigate a fabrication-demanding straddling regime of qubit frequencies, and even then the gate speed is typically well over 100 ns~\cite{sheldon_procedure_2016, kandala_demonstration_2021}, 
limited by leakage error under strong drives in a crowded spectrum. 

\begin{figure*}[btp]
    \centering
    \includegraphics[width=18.6cm]{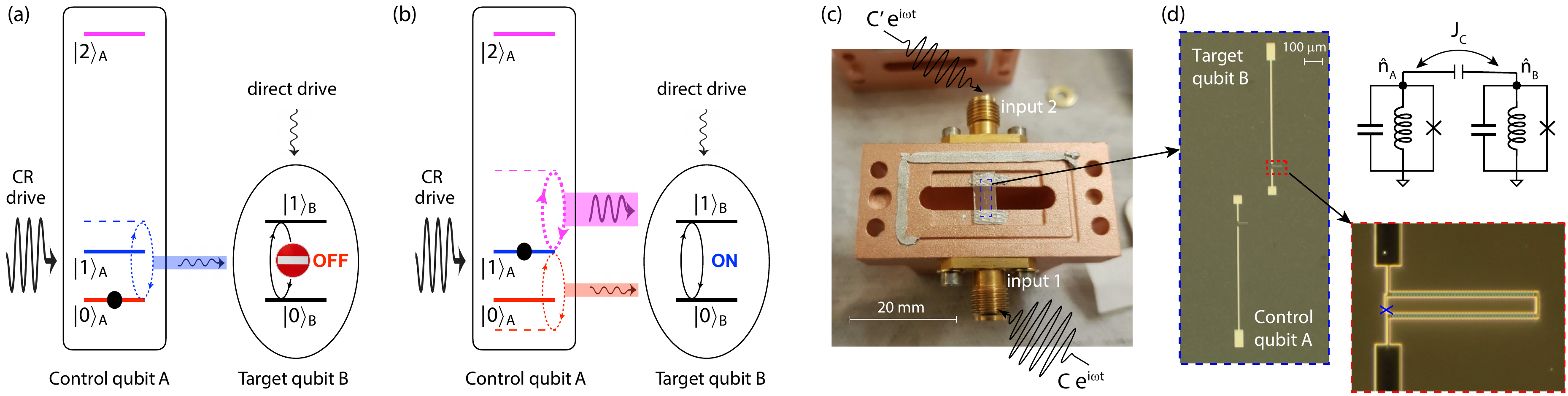}
    \caption{(a, b) The mechanism of the fluxonium cross-resonance effect, illustrated in the relevant bare energy level diagram of two fluxonium qubits with computational transition frequencies $\omega_A<\omega_B$ at half flux quantum.   A large CR drive and a small direct drive at $\omega_B$ are applied simultaneously to the two qubits.  (a) The OFF state of the fluxonium switch: When $A$ is in $\ket{0}$, the relatively weak $\ket{0}$-$\ket{1}$-mediated transmission of the CR drive (blue) is canceled by the direct drive. (b) The ON state of the fluxonium switch: When $A$ is in $\ket{1}$, the stronger $\ket{1}$-$\ket{2}$-mediated transmission of the CR drive (magenta) is activated.  The $\ket{0}$-$\ket{1}$-mediated drive also flips its sign (red) and adds constructively to the total CR effect. 
    Note that the $\ket{0}$-$\ket{2}$ transition is forbidden at half flux.   
    (c) Photo image of half of the cavity with two input ports, where the sapphire chip with two fluxonium qubits is located in the middle. (d) Photo image of the two fluxonium qubits, each made of an extended dipole antenna/capacitor, a junction chain superinductor (see the zoomed-in panel), and a small Josephson junction (at the position marked by a blue cross). The dipole of Qubit $A$ is extended towards Input Port 1 and hence stronger coupled to it, and vice versa for Qubit $B$.  The mutual capacitance of the antennas provides the capacitive coupling between the fluxoniums as in the effective circuit diagram.
    }
    \label{fig:0}
\end{figure*}

In this work, we experimentally demonstrate a  cross-resonance gate between a pair of capacitively-coupled fluxonium qubits. 
Following the proposal of the selective-darkening CR scheme~\cite{nesterov_controlled-not_2022}, our gate realizes a direct \textsc{CNOT} operation using vector compensation of two simultaneous drives applied to a 3D cavity.  We report \textsc{CNOT} gate fidelity above 99.4\% from interleaved randomized benchmarking with a gate time of 70 ns in a $ZZ$-cancelled two-qubit device ($\mathcal{F}=99.49(6)\%$ averaged over 7 hours, $\mathcal{F}=99.42(6)\%$ averaged over 4 separate trials spanning 2 months).  We further investigate gate performance at various drive power, and obtain high-fidelity ($\mathcal{F}>99\%$) \textsc{CNOT} gate as fast as 54 ns and with pulse ramp time as short as 2 ns.
The current gate performance is limited by an extraordinary microwave reflection problem of our drive lines (mitigated by a cancellation technique that we will discuss) and lower-than-expected fluxonium coherence times in our setup, which can be substantially improved in the near future.  \\


\textit{Fluxonium as a quantum switch} -- 
The working principle of the fluxonium CR gate can be conceptually illustrated using the language of virtual-state transitions~\cite{tripathi_operation_2019} in the bare single-qubit level diagram in Fig.~\ref{fig:0}.  We consider two capacitively coupled fluxonium qubits, where the low-frequency control qubit $A$ is driven at a moderately higher frequency ($\omega_B/2\pi\approx1$ GHz in our experiment) of the target qubit $B$.  The Hamiltonian is: 
\begin{equation}\label{Hamiltonian-two-qubit}
\hat{H} = \hat{H}_{A} + \hat{H}_B + J_C \hat{n}_A \hat{n}_B + \hat{H}_{\rm cr} 
+ \hat{H}_{\rm dr} 
\end{equation}
where the individual bare fluxonium Hamiltonian
($\alpha=A, B$)
\begin{equation}\label{Hamiltonian-fluxonium}
 \hat{H}_{\alpha} = 4E_{C,\alpha} \hat{n}_\alpha^2 + \frac 12 E_{L,\alpha} \hat{\varphi}_\alpha^2 - E_{J,\alpha} \cos(\hat{\varphi}_\alpha - \phi^{\rm ext}_{\alpha})\, ,
\end{equation}
is written in terms of the dimensionless flux ($\hat{\varphi}_\alpha$) and charge ($\hat{n}_\alpha$) operators and the charging ($E_{C,\alpha}$), inductive ($E_{L,\alpha}$), and Josephson ($E_{J,\alpha}$) energies. The capacitive interaction between the two fluxoniums is described by the coupling energy $J_C$.
The last two terms in Eq.~\eqref{Hamiltonian-two-qubit} describe the aforementioned cross-resonant drive,  
$\hat{H}_{\rm cr}=\varepsilon_A \hat{n}_A\cos{\omega_Bt}$, and a small additional drive applied directly to Qubit $B$, 
$\hat{H}_{\rm dr}=\varepsilon_B\hat{n}_B\cos{\omega_B t}$.

\begin{table*}[btp]
    \centering
    \begin{tabular}{cccccccccccc}
    \hline\hline\\[-1.9ex]
        Qubit ($\alpha$)
         & $\frac{E_L}{h}$ (GHz) & $\frac{E_C}{h}$ (GHz) & $\frac{E_J}{h}$ (GHz) & $\frac{\omega_{\alpha}}{2\pi}$ (GHz) & $\frac{\omega_{\alpha,12}}{2\pi}$ (GHz) & $\frac{J_c}{h}$ (GHz) 
         & 
         $|\langle 0 |\hat{n}_\alpha |1\rangle |$ & 
         $|\langle 1 |\hat{n}_\alpha |2\rangle |$ & 
         $T_1$ ($\mu$s) & $T^*_2$ ($\mu$s) & $T_{2E}$ ($\mu$s)
         \\[0.5ex]
         \hline\\[-2ex]
         $A$ & 0.78  & 1.18  & 4.03  & 0.5552  & 3.610\,/\,3.691 & \multirow{2}{*}{0.28} 
         & 0.13 & 0.55 & 52.0 - 60.0 & 14.0 - 15.5 & 22.0 - 24.0 
         \\
         $B$ &1.42  & 1.13  & 4.34  & 1.0045  & 3.719\,/\,3.796 & 
         & 0.20 & 0.59 & 17.0 - 33.0 & 5.5 - 7.0 & 13.0 - 16.5
         \\
         \hline\hline
    \end{tabular}
    \caption{Device parameters.  The qubit frequencies and the (range of daily-averaged) coherence times are measured at the operating flux point of the experiment.  
    The circuit Hamiltonian parameters and transition matrix elements are extracted from fitting the qubit spectroscopy data.}
    \label{Table-params}
\end{table*}

The highly anharmonic control qubit $A$ functions like a microwave switch due to its very state-dependent response to the off-resonance CR drive.  When the control is in $\ket{0}_A$, Qubit $B$ receives an effective drive field mediated by the virtually-excited $\ket{0}_A$-$\ket{1}_A$ transition (Fig.~\ref{fig:0}(a)), 
with a resonant Rabi rate:
\begin{equation}\label{TE0}
\Omega_0\approx\varepsilon_A\bigg[i\frac{J_C}{\hbar}\frac{\langle 0|\hat{n}_A|1\rangle^2}{\omega_B-\omega_A}\bigg]\langle 0|\hat{n}_B|1\rangle
\end{equation}
where the square bracket part can be understood as a ``transmission factor" of the CR drive, while the matrix element $\langle 0|\hat{n}_B|1\rangle$ 
factors in the response function of $B$ to any electric drive field it receives.  The transmission factor flips sign when Qubit $A$ is excited to $\ket{1}_A$, giving the original cross-resonance ZX Hamiltonian for the ideal spin-1/2 system~\cite{rigetti_fully_2010}.  This ZX effect is suppressed by the small charge matrix element of the low-frequency fluxonium transitions. However, the $\ket{1}_A$ state also opens a
stronger pathway of transmission via the virtual non-computational $\ket{1}_A$-$\ket{2}_A$ transition (Fig.~\ref{fig:0}(b)).  Even though this transition is several GHz detuned, $\omega_{A,12}\gg\omega_A,\omega_B$, its presence is felt strongly since the process doubly benefits from its much larger matrix element $\langle 1|\hat{n}_A|2\rangle$, giving: 
\begin{equation}\label{TE1}
\Omega_1\approx\varepsilon_A\big(-i\frac{J_C}{\hbar}\big)
\bigg[\frac{\langle 0|\hat{n}_A|1\rangle^2}{\omega_B-\omega_A}
+\frac{\langle 1|\hat{n}_A|2\rangle^2}{\omega_{A,12}-\omega_B}\bigg]
\langle 0|\hat{n}_B|1\rangle\, .
\end{equation}
This extra term describes an enhanced CR effect beyond the spin-1/2 system, which is known to provide up to a factor of 2 boost for the transmon CR rate in the straddling regime but requires precise frequency placements and may easily cause leakage error~\cite{tripathi_operation_2019}.  The situation for fluxonium is fundamentally different: This additional transmission factor contains no small parameters related to either qubit frequencies
and hence can be the dominant enabler of fast gates for low-frequency qubits.  It leads to a conditional rotation of Qubit $B$, and by applying a small compensation drive directly on $B$ ($\hat{H}_{dr}$) to fully cancel out its dynamics in the OFF state (known as selective darkening~\cite{de_groot_selective_2010}), a direct \textsc{CNOT} gate can be realized.  In this discussion, we have omitted the (non-negligible) $\ket{0}$-$\ket{3}$ contribution and non-RWA-like hybridizations; a more rigorous calculation has been carried out in Ref.~\cite{nesterov_controlled-not_2022}. 
Crucially, in contrast to previous microwave-based two-qubit gates that necessarily populate the non-computational states~\cite{ficheux_fast_2021, xiong_arbitrary_2022}, here the occupation probability of the $\ket{2}_A$ state is minimal since it scales inverse-quadratically with the multi-GHz drive detuning (while $\Omega_1$ scales inverse linearly).
\\

\textit{Device setup} -- 
While the local drives required in the CR scheme may fit more naturally with planar architectures, here we use a convenient 3D circuit QED design~\cite{paik_observation_2011} to carry out a proof-of-principle demonstration. Two capacitively coupled fluxoniums are fabricated on a sapphire substrate and enclosed in a copper cavity, and the Hamiltonian parameters extracted from the two-tone spectroscopy are listed in Table I.  Compared to a similar device in Ref.~\cite{ficheux_fast_2021}, our cavity has two drives ports placed on the opposite sides of the sapphire chip.  Each port has stronger coupling to one of the qubits due to the asymmetric layout of the fluxoniums on chip, which gives us spatial selectivity to apply drives to the two-fluxonium circuit (Fig.~\ref{fig:0}(c, d)).

A large external superconducting coil is used to apply a static global magnetic field to the circuit.  We carry out our experiment at a fixed bias field giving $\phi^{\rm ext}_{A}/2\pi=0.5005$, $\phi^{\rm ext}_{B}/2\pi=0.4993$, when both qubits are within 
0.7 MHz from their exact half-flux ``sweet spots". The coherence times for both fluxoniums at this operating point are noted in Table I, which are primarily not limited by flux noise (i.e.~insensitive to the exact choice of external flux in the vicinity of this operating point). We attribute the subpar $T_1$ times to a combination of higher dielectric loss and insufficient infrared shielding compared to the current of the art~\cite{somoroff_millisecond_2021}.  Both low-frequency noise and cavity photon shot noise contribute to the low $T_2$ times. 

The relatively strong capacitive coupling between the two fluxoniums results in a static $ZZ$ interaction $\xi^0_{ZZ}/2\pi
= 0.9$ MHz.  While $\xi^0_{ZZ}$ does not limit the fidelity of the direct CNOT gate thanks to its selective-darkening construction~\cite{nesterov_controlled-not_2022},  it poses cross-talk challenges in scaling up multi-qubit systems.  Throughout our experiment, we apply an off-resonant continuous microwave tone (at 3.850 GHz, 54 MHz detuned from the $\ket{11}$-$\ket{21}$ transition) at all times to cancel the $ZZ$ interaction by differential ac Stark shift (giving residual $\xi_{ZZ}/2\pi<20$ kHz) ~\cite{xiong_arbitrary_2022}.  This $ZZ$ cancellation pump results in an estimated 2\% $\ket{11}$-$\ket{21}$ hybridization, and the impact to qubit coherence is experimentally very minimal.  Future studies may further employ multi-path or tunable couplers to suppress $\xi_{ZZ}$~\cite{nguyen_scalable_2022, moskalenko_tunable_2021} if needed.  For the rest of the paper, we will work exclusively with the dressed eigenstates under the pump tone, which are labeled $\ket{00}, \ket{01}, \ket{10}, \ket{11}$ and form the computational basis of the two-qubit system. 

We initialize our qubits in the $\ket{00}$ state with an estimated fidelity of about 94\% using a cavity-sideband cooling procedure.  This protocol effectively dumps the entropy of the low-frequency qubits to the cold bath of the high-frequency cavity over a duration of 15 $\mu$s.  Due to the lack of a quantum-limited parametric amplifier in our experiment, the joint two-qubit state is measured by analyzing the averaged readout transmission signal with different combinations of qubit pre-rotations preceding the measurement~\cite{filipp_two-qubit_2009}. See Supplementary Material~\cite{supp} for more details of state initialization and readout.\\

\textit{Characterization of cross-resonance dynamics} -- 
To realize controlled operation of Qubit $B$, we apply microwave drives at $\omega_B$ simultaneously to both physical input ports. The two drives are produced by separate IQ modulation of the same local oscillator source. Therefore they are phase locked from each other but have independently controllable complex amplitude $C$ and $C'$.  The CR drive amplitude $\varepsilon_A$ and direct drive amplitude $\varepsilon_B$ locally incident on the two qubits are linearly related to $C$ and $C'$ by a complex-valued 2x2 ``classical cross-talk" matrix which we do not need to explicitly characterize.  We simply find a complex ratio $\eta=C'/C$ experimentally to darken the $\ket{00}-\ket{01}$ transition, 
so that the combined effect of $\hat H_{cr}$ and $\hat H_{dr}$ gives a rotating-frame effective drive Hamiltonian in a block-diagonal form (written in the form of $A\otimes B$):
\begin{equation}
\hat{H}_\text{drive}= \frac{\Omega}{2} \ket{1}\bra{1}\otimes \sigma_x + \frac{\Delta_s}{2} \sigma_z \otimes \mathbb{I}
\label{eq:HCX}
\end{equation}
where the first term denotes a conditional X rotation of the target qubit (equivalent to $\frac{\Omega}{4} (\mathbb{I}-\sigma_z)\otimes \sigma_x$) and the second term represents an ac Stark shift $\Delta_s$ on the control qubit.

\begin{figure*}[tbp]
    \centering
    \includegraphics[width=18.5cm]{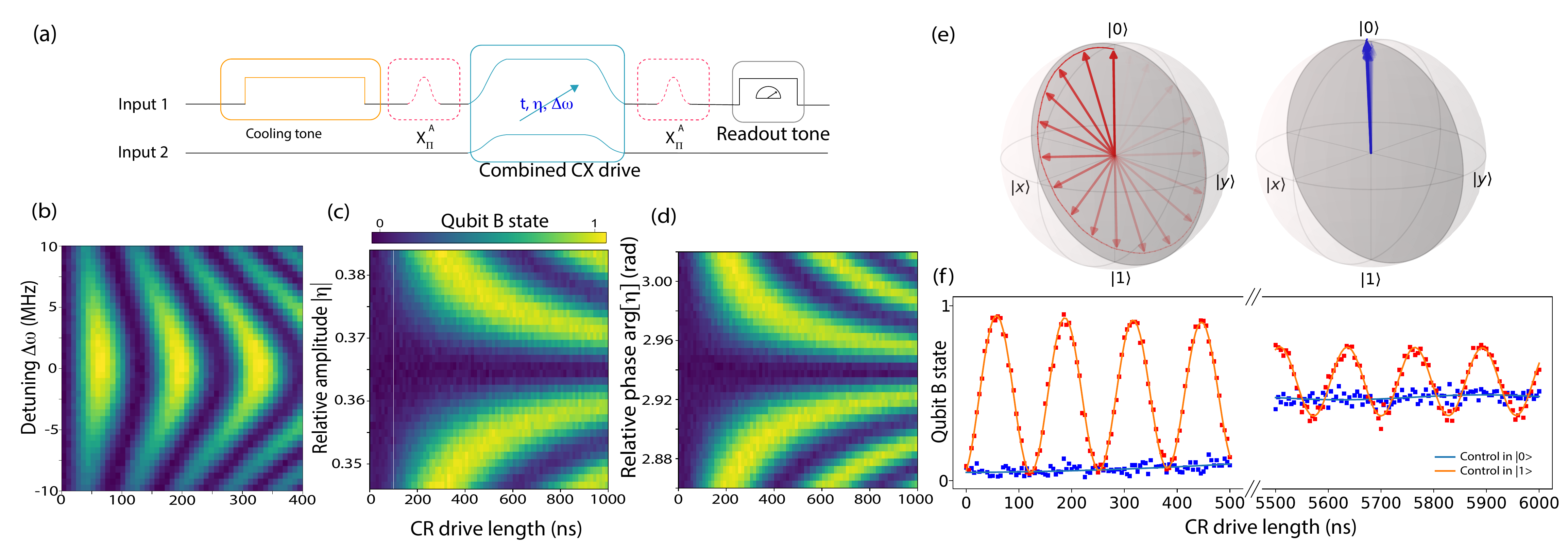}
    \caption{(a) Pulse sequence for measuring the controlled driven dynamics of Qubit $B$. Here the first $X_\pi^A$ pulse is applied (skipped) to set the Control = \textsc{ON} (OFF) state, and the second $X_\pi^A$ pulse is applied (skipped) to reset $A$ to $\ket{0}$ for consistent readout contrast. (b) Rabi oscillation of $B$ at the ON state as a function of time and drive detuning under the optimized complex drive ratio $\eta$ for CR gate.  (c, d) Dynamics of $B$ at the OFF state when $\eta$ deviates in amplitude (c) and phase (d) from the perfect darkening condition.  $\eta$ is in arbitrary unit as it includes the difference in attenuation and electrical delay of the two drive lines.  (e, f) Conditional Rabi oscillation of $B$ under resonant CR drive, comparing its dynamics at the ON (red) and OFF (blue) states.  (e) shows the tomographically reconstructed Bloch sphere trajectory of Qubit $B$ over the first oscillation period of (f).  The loss of oscillation contrast over long time is consistent with qubit coherence times. All data is taken with a combined drive amplitude that gives an estimated effective CR drive strength $\epsilon_A/\mel{0}{\hat{n}_B}{1}$ of 90 MHz (by comparing to the Rabi rate of $B$ under the drive of individual input ports).
    }
    \label{fig:2}
\end{figure*}

To calibrate the complex drive ratio $\eta$, we initialize $A$ in $\ket{0}$ and null the Rabi oscillation of $B$ by sweeping the relative amplitude ($|\eta|$) and the relative phase (arg$[\eta]$) of the drives.  After fixing $\eta$, we still retain the freedom of choosing $C$, or the overall amplitude scale and common phase of the two drives, which endows full control of the $\ket{10}-\ket{11}$ subspace.  When $A$ is initialized in $\ket{1}$, $B$ displays the prototypical Rabi dynamics, as shown in Fig.~\ref{fig:2}(b) over a range of drive detunings
for a given drive power.  We can further reconstruct the Bloch-sphere trajectory of qubit $B$ under the CR drive by performing single-qubit state tomography at different times for control qubit in $\ket{0}$ and $\ket{1}$ respectively (Fig.~\ref{fig:2}(c)), which demonstrates that the desired Hamiltonian Eq.~(\ref{eq:HCX}) has been realized.  
We can also vary the conditional Rabi rate $\Omega$ as a function of time using a drive envelope $C(t)$, and any envelope that gives an integrated rotation angle $\int \Omega(t) dt=\pi$ would yield a CNOT-equivalent controlled $X_\pi$ rotation. \\

\textit{Single-qubit control and pulse reflection correction} --
We use a similar calibration procedure to obtain unconditional single qubit rotations.  Since microwave applied to either input ports contributes to both $\hat{H}_\text{dr}$ and $\hat{H}_\text{cr}$ but at different ratio, to realize clean single qubit rotation of $B$ independent of $A$, we apply microwave drives at $\omega_B$ to both ports simultaneously with a complex drive ratio $\eta'$ to null the $\sigma_z\otimes\sigma_x$ term of the drive Hamiltonian~\cite{supp}.  Because the CR effect from the higher-frequency $B$ to the lower-frequency $A$ is weaker for our device parameters, in practice we did not find it necessary to pursue the compensated drive scheme for $A$ and  simply used Port 1 to drive it. 

Fast high-fidelity gates in general require impedance-matched transmission lines free of standing waves.  As gate fidelity improves, impedance mismatch can start to cause appreciable harm, whose characterization and mitigation may require careful studies.  Accidentally, we had to carry out the experiment under a challenging condition with extreme reflection problems in our drive lines.  
Each control pulse bounces off the cavity multiple times, among which the most pronounced secondary impact arrives about 20 ns later than the original pulse and carries $\sim$35\% of the original amplitude.  
We developed an ad hoc procedure to calibrate the timing and complex amplitude of the reflected pulses, 
which extends the method in Ref.~\cite{gustavsson_improving_2013} to both I and Q quadratures~\cite{supp}.  By programming waveform pre-distortion to all control pulses to cancel the reflections throughout our experiment, we 
improved single-qubit gates from a completely meaningless mess to average fidelity of $\mathcal{F}\gtrsim99.8\%$ for $A$ and $\mathcal{F}\gtrsim99.7\%$ for $B$ as measured by simultaneous randomized benchmarking~\cite{supp}.  For both qubits, each single-qubit gate uses a 16 ns pulse with a Gaussian envelope ($\sigma=4$ ns). The fidelity remains lower than the qubit coherence limit, which we attribute to the imperfect waveform correction. \\



\textit{Calibration and characterization of the $CX_\pi$ gate} --
Under the CR drive Hamiltonian Eq.~(\ref{eq:HCX}), the dynamics in Fig.~\ref{fig:2}(c) provides a coarsely tuned controlled-$X_\pi$ rotation ($\textsc{CX}_\pi$) after a half Rabi period. 
To bring up a high-fidelity two-qubit gate, we fine tune the CR drive parameters and account for extra single-qubit phases ($\theta_A$ and $\theta_B$) with rotating frame updates (known as virtual Z rotations~\cite{mckay_efficient_2017}) to realize  
the $\textsc{CX}_\pi$ gate: 
\begin{equation}
 \begin{pmatrix}
  1 & 0 & 0 & 0 \\
  0 & e^{i\theta_{B}} & 0 & 0 \\
  0 & 0 & 0 & -ie^{i\theta_{A}} \\
  0 & 0 & -ie^{i\theta_{A}} & 0 
 \end{pmatrix}\
 \rightarrow
 \textsc{CX}_\pi = 
 \begin{pmatrix}
  1 & 0 & 0 & 0 \\
  0 & 1 & 0 & 0 \\
  0 & 0 & 0 & -i \\
  0 & 0 & -i & 0 
 \end{pmatrix}\,
\end{equation}
where the top-left and bottom-right blocks correspond to the $\ket{0}_A$ and $\ket{1}_A$ subspaces respectively.  
The $\textsc{CX}_\pi$ gate is connected to the textbook \textsc{CNOT} gate by a \textsc{S} gate on $A$, which can be trivially absorbed in $\theta_A$ at no additional cost.  We targeted $\textsc{CX}_\pi$ instead of \textsc{CNOT} in our calibration and verification solely out of our custom software convention.  For all purposes our result can be viewed as applicable for a CNOT gate.

\begin{figure}[tbp]
    \centering
    \includegraphics[width=8.0cm]{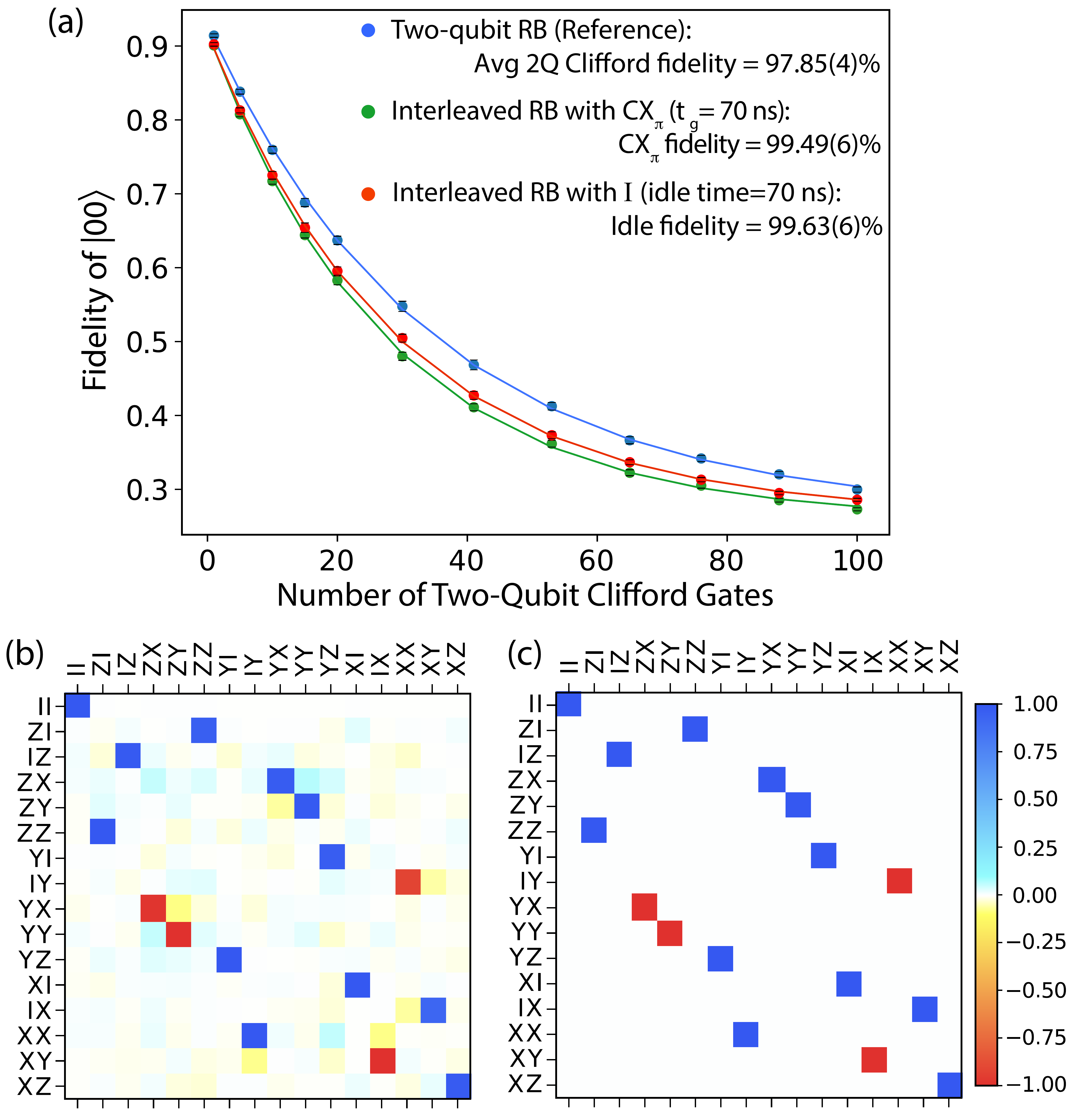}
    \caption{(a) Interleaved randomized benchmarking (IRB) of the CR gate calibrated for a total gate length of 70 ns (including two 6-ns Gaussian ramp edges).  The plot is averaged over 60 randomly generated Clifford sequences per data point over a continuous run of 7 hours with periodic automatic calibrations.  The reference error per 2-qubit Clifford (2.15\%) is consistent with the average number of physical single-qubit gates (6.5) and two-qubit gates (1.5) per Clifford and their fidelity.  Also shown are IRB of a 70 ns idling gate, which provides an upper bound of the decoherence error.  
    (b) The reconstructed process matrix of our CR gate by quantum process tomography (QPT).  The QPT is performed by applying the CR gate to 36 different initial states, each followed by state tomography which takes 29 different pre-rotation configurations before the joint readout. 
    (c) The ideal process matrix of $\textsc{CX}_\pi$ for comparison. }
    \label{fig:RBQPT}
\end{figure}

\begin{figure}[tbp]
    \centering
    \includegraphics[width=7cm]{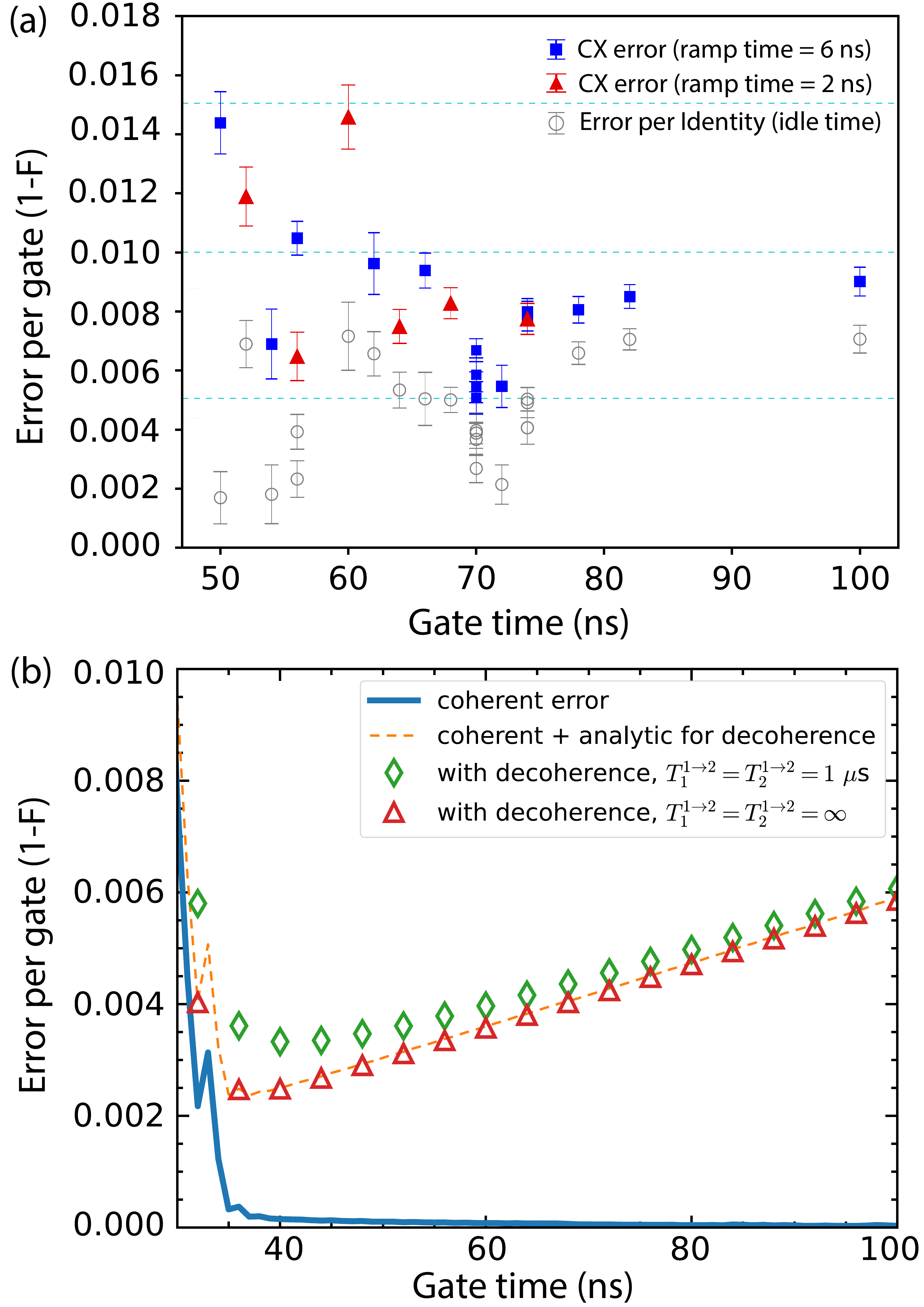}
    \caption{(a) The two-qubit $\textsc{CX}_\pi$ gate error extracted from interleaved randomized benchmarking (IRB) versus gate time for rounded square pulses with ramp times of 6 ns (blue) and 2 ns (red).  Also plotted for comparison are idling error from IRB (grey). (b) Numerical simulation results of coherent error per gate (solid blue curve), total error per gate including decoherence of computational states (red triangles), and total error per gate further including decoherence of the non-computational $\ket{2}$ states.  The decoherence rates of the $\ket{2}$ states are exaggerated ($T_1=T_2=1$ $\mu$s) to demonstrate that their effect is very minimal.  The decoherence rates of the computational states in the simulation use $T_1$ and $T_{2E}$ values in Table I, whose contribution to gate error is in good agreement with the analytical estimate of $t_g/T_{\rm err}$ (orange dashed line).  
    The simulation follows the numerical procedure in Ref.~\cite{nesterov_controlled-not_2022} and is based on the fluxonium parameters from the experiment and considers rounded square pulses with ramp time of 6 ns.}
    \label{fig:GLSim}
\end{figure}

Our $\textsc{CX}_\pi$ gate uses a rounded square pulse envelope typically with 6 ns half Gaussian ($\sigma=3$ ns) rising and falling edges.  At a fixed gate time, we iterate through dedicated subroutines sensitive to specific control errors to calibrate against 7 parameters: The relative amplitude and phase (complex $\eta$) of the CR drive to ensure selective darkening of the $\ket{00}-\ket{01}$ transition, the common amplitude and phase (complex $C$) and the detuning of the CR drive to ensure a precise $X_\pi$ rotation in the $\ket{10}-\ket{11}$ subspace, the relatively large single-qubit phase $\theta_A$ due to the ac Stark shift ($\sigma_Z\otimes\mathbb{I}$ term in Eq.~(\ref{eq:HCX})), and a small phase $\theta_B$ possibly due to a spurious $\mathbb{I}\otimes\sigma_Z$ Hamiltonian from a higher-order CR effect~\cite{nesterov_controlled-not_2022}.  These 7 parameters cover all possible control errors of $\textsc{CX}_\pi$ within its block diagonal structure.  We further monitor possible spurious rotations of $A$ (i.e.~leakage drive on $A$) to ensure the process is block diagonal.  See Supplementary Material~\cite{supp} for details of the calibration procedure. A similar routine has been described in a recent report of CR gate in transmons~\cite{wei_quantum_2021}.

We use quantum process tomography (QPT) and interleaved randomized benchmarking (IRB) to characterize the performance of the CR gate.  Fig.~\ref{fig:RBQPT} shows the result for our calibrated CR gate at an optimal gate length of 70 ns. Our process tomography follows the procedure outlined in Ref.~\cite{chow_universal_2012}. The reconstructed process matrix is in excellent agreement with the ideal $\textsc{CX}_\pi$ gate with no outstanding spurious non-zero elements, qualitatively confirming the performance of the gate.  The extracted process fidelity is 99.1\% although the QPT fidelity is known to be sensitive to the underlying model of rescaling state preparation and measurement (SPAM) infidelity.  The IRB provides a more SPAM-agnostic validation of the CR gate fidelity.  Following the IRB procedure~\cite{corcoles_process_2013, barends_superconducting_2014}, we compare the sequence fidelity versus sequence length for random two-qubit Clifford gates (reference) and those interleaved with additional $\textsc{CX}_\pi$ gates, giving the $\textsc{CX}_\pi$ gate fidelity $\mathcal{F}=$ 99.49(6)\% (Fig.~\ref{fig:RBQPT}(a)).\\


\textit{Gate time and pulse shape} --
We calibrated and benchmarked the $\textsc{CX}_\pi$ gate at different gate times in the 50-100 ns range, and the IRB fidelity is shown in Fig.~\ref{fig:GLSim}(a).  We achieved CR gate fidelity above 99\% over a broad range of gate times, and the fastest well-performing gate takes only 54 ns ($\mathcal{F}\approx99.3\%$).  This is a major speed-up from the transmon CR gates, which has been typically in the range of 150 ns or longer~\cite{sheldon_procedure_2016, kandala_demonstration_2021, patterson_calibration_2019} and only very recently reaching a record of 90 ns~\cite{wei_quantum_2021}.
We also benchmarked the fidelity of idling gates with different (idling) time using IRB, which has been used as a sensitive noise spectrometer to probe qubit decoherence on the same time scale as fast gate operations~\cite{omalley_qubit_2015}.  However, we observed unusual non-monotonic behavior in idling fidelity which cannot be explained by any Markovian noise models.  While part of the fluctuations may be attributed to qubit coherence time fluctuations over the 2 months of data acquisition, we believe this observation is a clear evidence that the residual pulse reflection problem remains a major contributing factor in our gate error and is responsible for the unexpected peaks and valleys of the CR gate fidelity with respect to gate time.  

The optimal choice of the gate time in principle is determined by the trade-off between the coherent control error at the short-time limit and the incoherent error at the long-time limit.  Numerical simulation of the CR gate performance~\cite{nesterov_controlled-not_2022} using our device parameters illustrates this trade-off clearly (Fig.~\ref{fig:GLSim}(b)): the unitary error decreases fast with increasing gate time $t_g$, while the incoherent error increases linearly in time as $t_g/T_{\rm err}$, where $T_{\rm err}^{-1} = (T_{1,A}^{-1} + T_{1,B}^{-1} +2T_{2E,A}^{-1} +2T_{2E,B}^{-1})/5
$~\cite{tripathi_operation_2019}. 
The simulation suggests further room to speed up the gate to the 40 ns range for optimal performance in a system free of reflection problems. 

Another major advantage of highly anharmonic qubits like the fluxonium is the relaxed requirements on pulse shaping.  The transmons CR gates often require careful pulse shaping and slow ramp times (at least 15 ns~\cite{kandala_demonstration_2021, wei_quantum_2021}) to avoid driving unwanted transitions.  The only constraint in pulse shapes in our fluxonium CR gate is to avoid off-resonant excitation of the Qubit $A$ (hence limited by qubit-qubit detuning), and once again the small matrix element of the low-frequency Qubit $A$ works in our favor.  We further tested reducing the pulse ramp from 6 ns to 2 ns, which is faster than the bandwidth of our arbitrary waveform generator (350 MHz, with 1 ns digital resolution), and observed no apparent impact in CR gate fidelity (Fig.~\ref{fig:GLSim}).\\

\textit{Outlook}--
Compared to flux-controlled two-qubit gates, the all-microwave cross-resonance gate has been historically considered a gate that sacrifices speed for simplicity and noise protection.  This fundamental trade-off remains front and center in today's competition to scale with transmon-based architectures.  Armed with large anharmonicity and favorable selection rules (i.e.~hierarchy of matrix elements), the fluxonium qubits have opened a new dimension for engineering the CR effect and are poised to make the best of both worlds.  
Among the first generation of two-fluxonium gates emerged recently, our CR gate is on par in speed with the flux-controlled gates~\cite{bao_fluxonium_2021, moskalenko_high_2022} without the need of extra coupler elements or fast flux biasing. 

Most importantly, the fluxonium CR gate is expected to well preserve the coherence properties of the computational subspace at half-flux.  Although the lingering pulse reflection problem prevented us from metrology of possible control errors much below the reported gate infidelity, our best gate (at 70 ns) performs very close to the free-evolution coherence limit (see Fig.~\ref{fig:GLSim}(b)), showing no hidden obstacles (such as drive-induced instability or decoherence) at least at the level of $1\times10^{-3}$.  Since our coherence times have more than one order of magnitude of room for improvement, and numerically predicted control error is easily below $10^{-4}$~\cite{nesterov_controlled-not_2022}, one can be optimistic of rapid fidelity improvement from the current level.  Assuming the modest capacitive quality factor as in Ref.~\cite{somoroff_millisecond_2021} and that $T_{2E}$ can catch up to $T_1$ with sufficient thermalization~\cite{nguyen_high-coherence_2019, somoroff_millisecond_2021}, the two qubits at their current working frequency would have coherence times of 400 and 200 $\mu$s respectively, projecting a coherence-limited gate error rate of $2\times10^{-4}$ at 50 ns gate time. 
Assuming more transmon-like quality factor for fluxonium (as reported in Ref.~\cite{pop_coherent_2014}) 
may potentially bring another $\sim5\times$ improvement.  We further note that the Hamiltonian parameters of our fluxonium qubits are by no means optimal, and there is a very large design space to explore optimization strategies.  

The selective-darkening CR effect of fluxonium qubits can be more generally viewed as a quantum switch for a broad bandwidth of external control fields. 
It provides the means for not just a single two-qubit gate (\textsc{CNOT}), but also a broader class of controlled unitary operations.  
We may trivially modify the phase of the CR drives to implement a $\textsc{CY}_\pi$ gate in addition to the $\textsc{CX}_\pi$ with the same quality, and it is also straightforward to perform any combination of controlled X- or Y-rotations of arbitrary angles.  One can also detune the CR drive to obtain controlled Z-rotations as was done for transmons~\cite{wei_quantum_2021, mitchell_hardware-efficient_2021}, hence allowing for highly efficient controlled arbitrary single-qubit operations. Such controlled operation may be further extended to manipulating multi-mode bosonic or multi-qubit systems.  
Since controlled unitary is one of the most important building blocks for many quantum algorithms and subroutines (e.g.~phase estimation), native parameterized controlled unitary operations will likely bring substantial benefit to NISQ applications.\\

\begin{acknowledgments}
\textit{Acknowledgments}-- 
We thank Quentin Ficheux, Long Nguyen, Joseph Bardin for helpful discussions. This research was supported by the ARO-LPS HiPS program (No.~W911-NF-18-1-0146). V.E.M.~and M.G.V.~acknowledge the Faculty Research Award from Google and fruitful conversations with the members of the Google
Quantum AI team.  L.L.G.~acknowledges support from Google.  
\end{acknowledgments}

\bibliography{References}

\end{document}


\title{Supplemental Materials for\\
`Demonstration of the Two-Fluxonium Cross-Resonance Gate'
}
\author{Ebru Dogan}
\affiliation{Department of Physics, University of Massachusetts-Amherst, Amherst, MA, USA}
\author{Dario Rosenstock}
\affiliation{Department of Physics, University of Massachusetts-Amherst, Amherst, MA, USA}
\author{Lo\"ick Le Guevel}
\affiliation{Department of Physics, University of Massachusetts-Amherst, Amherst, MA, USA}
\affiliation{Department of Electrical and Computer Engineering, University of Massachusetts-Amherst, MA, USA}
\author{Haonan Xiong}
\affiliation{Department of Physics, Joint Quantum Institute, and Center for Nanophysics and Advanced Materials, University of Maryland, College Park, MD, USA}
\author{Raymond A. Mencia}
\affiliation{Department of Physics, Joint Quantum Institute, and Center for Nanophysics and Advanced Materials, University of Maryland, College Park, MD, USA}
\author{Aaron Somoroff}
\affiliation{Department of Physics, Joint Quantum Institute, and Center for Nanophysics and Advanced Materials, University of Maryland, College Park, MD, USA}
\author{Konstantin N.~Nesterov}
\affiliation{Department of Physics and Wisconsin Quantum Institute, University of Wisconsin–Madison, Madison, WI, USA}
\author{Maxim G.~Vavilov}
\affiliation{Department of Physics and Wisconsin Quantum Institute, University of Wisconsin–Madison, Madison, WI, USA}
\author{Vladimir E.~Manucharyan}
\affiliation{Department of Physics, Joint Quantum Institute, and Center for Nanophysics and Advanced Materials, University of Maryland, College Park, MD, USA}
\author{Chen Wang}
\email{wangc@umass.edu}
\affiliation{Department of Physics, University of Massachusetts-Amherst, Amherst, MA, USA}

\date{\today}

\maketitle

\section{Pulse predistortion via active reflection cancellation}

Within our current fridge setup, an impedance mismatch in the signal transmission chain severely impedes the integrity of the microwave pulses at the two qubit frequencies: Clean microwaves pulses played into our qubit input lines generate
several reflections hitting the qubit at a later time (on the order of 10's of ns) with different amplitude and phase, leading to severe deviation from the intended rotation.
More critically, the pulse reflections overlap with subsequent pulses resulting in unpredictable behavior of gate sequences such as contrast loss in power/time Rabi and strongly distorted chevron patterns (see Supplementary ~Fig.~\ref{suppfig:0}(a) and 1(c)). 
We attribute the higher-than-usual reflections to a non-cryogenic directional coupler placed at the mixing chamber with its coupled port connected to the cavity through a few eccosorb filters (see Supplementary~Fig.~\ref{suppfig:11}(b). The un-matched nature of the cavity inputs combined with the coupled port load deviating from the ideal 50-$\Omega$ value at cryogenic temperature and the absence of attenuation on the reflection path results in this critical situation. 

\begin{figure*}[htp]
    \centering
    \includegraphics{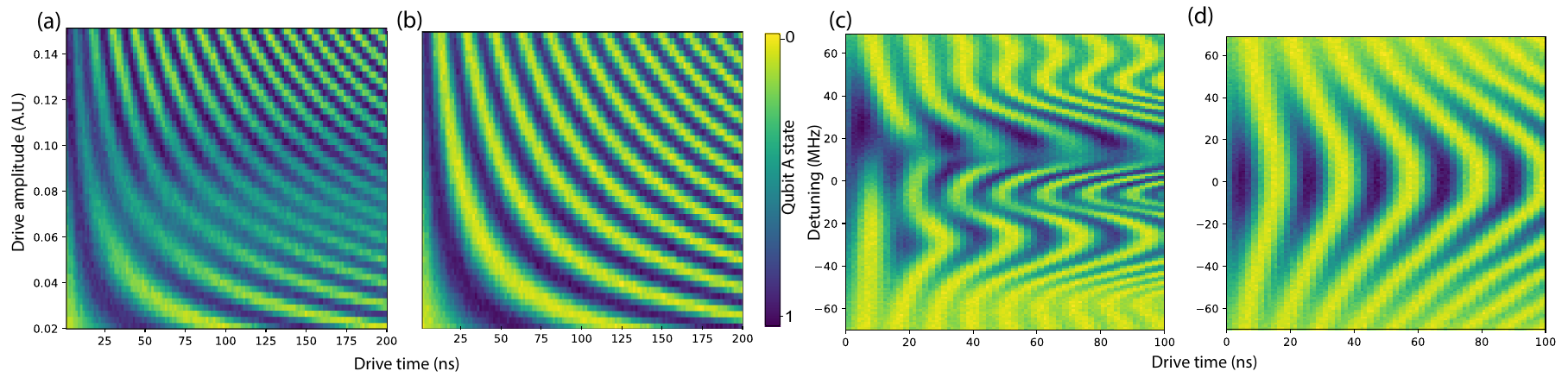}
    \caption{(a, b) Rabi oscillation of Qubit $A$ as a function of drive time and amplitude under a single-port drive without (a) and with (b) the predistortion: Problematic contrast loss is resolved and the correct oscillation frequency is recovered for (b).  (c, d) Rabi oscillation of Qubit $A$ as a function of drive time and drive detuning (the ``Chevron pattern") for a single-port drive without (c) and with the predistortion (d), driven at an amplitude corresponding to the $\pi$ pulse used throughout the experiment. The time axis does not include the ramp time of the pulse, which is 6 ns for each edge.  The reflection problem applies to both qubits and both drive lines: Similar traces are observed for all four combinations. }
    \label{suppfig:0}
\end{figure*}

\begin{figure*}[tbp]
    \centering
    \includegraphics{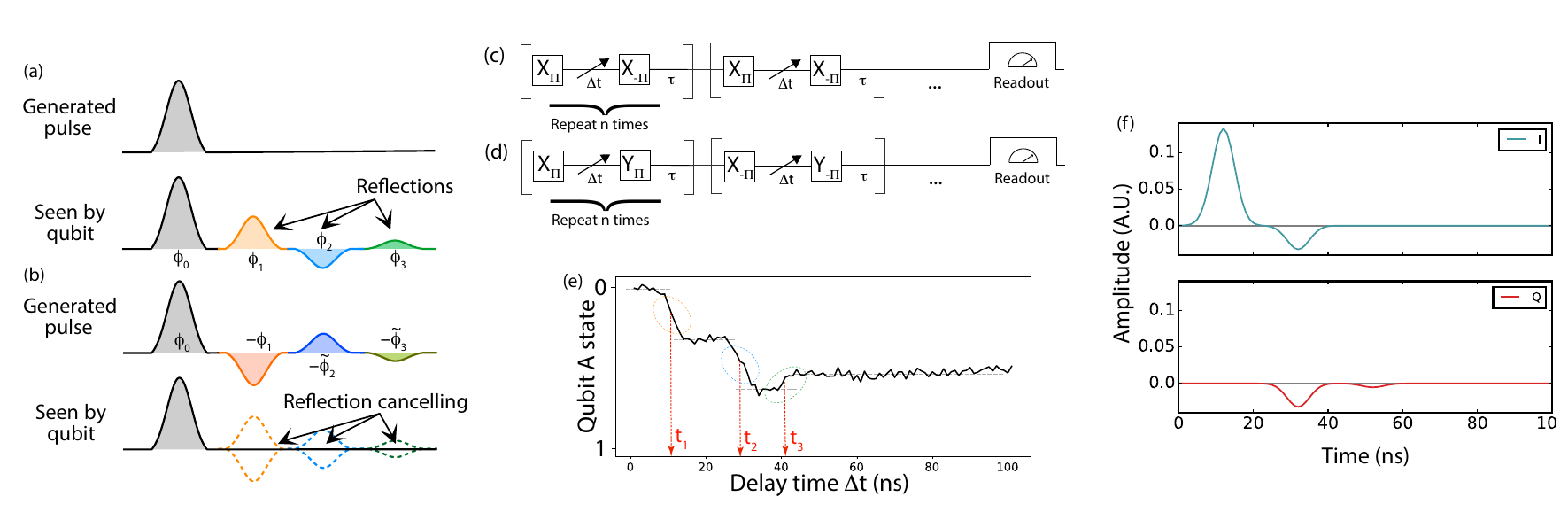}
    \caption{(a, b) Illustration of the generated pulse envelope and resulting envelope at the qubit port without any correction (a) and with predistortion (b). The phases for reflections in (b) are not the same as in (a) as each correction implemented creates its own reflections that changes the phase of the ones following it. (c, d) The pulse sequences for characterizing the quadrature-phase (c) and in-phase (d) components of reflections: We sweep over the delay time $\Delta t$ for a given number of repetitions n. $\tau$ is a time long enough (80 ns) for reflections from each set to not overlap with the next set or the readout. We start by n=1 and proceed by increasing up to n=7 for further amplification of the features. For the sequence in (d), the sign of the $\pi-$pulse changes for each iteration. (e) An example trace from the sequence from (c) (n=1) for characterizing the quadrature-phase reflection. Each jump that makes the qubit deviate from its previous state can be attributed to an additional reflection. The delay times with which these reflections follow the original pulse can be pinned down by summing the $X_{\pi}$ pulse length with times $t_{i}$ pointed out with red arrows in the figure. For both in-phase and quadrature-phase reflection characterizations, we use $X_{\pi}$ pulses as fast as 8 ns. (f) The set of envelopes that would need to be played from AWG I and Q channels in order to have a clean $X_{\pi}$ rotation for Qubit $A$. }
    \label{suppfig:1}
\end{figure*}

We mitigated this issue by applying a simple predistortion model to the qubit pulses: Our model consists of duplicating the ideal pulse into a train of delayed pulses whose timing, amplitudes, and phases are chosen to overlap and cancel the mismatch-induced reflections at the qubit port (Supplementary ~Fig.~\ref{suppfig:1}(a)).
All corrections are applied in the background and do not affect the effective qubit gate times.

\subsection{Reflection characterization}
Pulse distortions due to impedance mismatch issues are not unknown to the superconducting qubit community but mainly arise in fast flux lines driving a low-impedance inductor and the suggested pulse predistortion schemes mostly target the flux drive lines~\cite{jerger_situ_2019, rol_time-domain_2020}.  Our case of the microwave pulse reflection in our qubit drive lines is more relatable to an earlier work \cite{gustavsson_improving_2013} from which we make use of a pulse sequence characterizing the Q-quadrature reflection and extend it to our own model.

We start by defining a reflection with respect to the main pulse by three parameters: its delay time, relative amplitude and relative phase.
More conveniently in the qubit context, the reflections in the IQ plane are defined with the relative in-phase and relative quadrature-phase components. 
Equivalently, we can characterize these reflection parameters independently with relative amplitudes for in-phase and quadrature-phase components at their relevant delay times.  

The first step of the characterization is pinning down the delays of these reflections. An intuitive way of extracting them is by varying the delay between two following rotations: $X_{\pi}$ and $X_{-\pi}$: This is a sequence that isolates and accumulates the quadrature-phase component for the reflections while cancelling the in-phase component \cite{gustavsson_improving_2013}. For short delay times where the reflections for $X_{\pi}$ arrive after the pulse $X_{-\pi}$ with no overlaps, all reflections due to $X_{\pi}$ should be cancelled out by the reflections of $X_{-\pi}$: The measurement should result in ground state.
As the delay increases, reflections of $X_{\pi}$ start overlapping with the pulse $X_{-\pi}$. For sufficiently short pulses, the result is the step-like evolution of the qubit state where we can clearly see the effects of individual reflections (see Supp.~Fig.~\ref{suppfig:1}(e)): 
It is possible to identify each reflection from such a trace as each of the ``jumps" can be attributed to an additional reflection hitting the qubit. The points with the highest derivatives on these traces can be pinned down to determine the delay times for these reflections. 

We then use the same experiment with our targeted gate length and adjust the quadrature reflection component associated with each delay until the qubit state remains in ground for all delays, a sign of a good predistortion canceling the reflections at the qubit ports.
$X_{\pi},X_{-\pi}$ pairs are repeated up to 7 times, amplifying the deviation from ground, in a single measurement to finely tune the quadrature components (Supplemenary Fig.~\ref{suppfig:1}(c)).\\

The in-phase component is tuned in a similar measurement consisting of varying delay between $X_\pi$ and $Y_\pi$ rotations (Supplementary Fig.~\ref{suppfig:1}(d)).
The pulse sequence for the in-phase component characterization is sensitive to both in-phase and quadrature components: It is beneficial to have a reasonably well tuned cancellation of the quadrature component before moving on with the in-phase component. We then need to proceed with an iterative modification of both I and Q of each reflection, until both pulse sequences result in the designated final states with no significant deviations. 

Both qubits' drives require this protocol to be implemented, with the case of Qubit $B$ being more challenging than Qubit $A$ for being a combination of two individual drives. A sample envelope played from the AWG for an $X_{\pi}$ for the Qubit $A$ can be seen from Supplementary Fig.~\ref{suppfig:1}(f): Rather than a single gaussian on the I channel for a $\pi$ rotation around the $X$-axis, a set of two envelopes modified for reflection cancellations are played on both I and Q channels.

\begin{figure}[tbp]
    \centering
    \includegraphics[width=6cm]{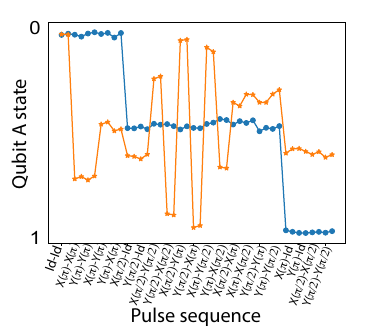}
    \caption{AllXY experiment\cite{reed_entanglement_2013} for the Qubit $A$ without (orange) and with (blue) the reflection cancellation scheme after the same set of tune up procedures. AllXY experiment consists of three sets of sequences: The first set of points land in the ground state, the second in the equator and the third in the excited state; pulse errors will show up as deviations from this configuration.}
    \label{suppfig:4}
\end{figure}

We depend on this predistortion scheme for having meaningful and reliable operations in our system. An AllXY experiment \cite{reed_entanglement_2013} as a visual indicator of the level of improvement can be seen from Supplementary Fig.~\ref{suppfig:4}.

\section{Calibration of initial state population with an entangling gate}

Without a single-shot readout, we take extra steps to calibrate the initial ground state population of the qubits, i.e.~to measure our state initialization fidelity.  This may be accomplished by making use of higher transitions: Comparing the amplitudes of $\ket{1}$-$\ket{2}$ oscillations with and without population inversion~\cite{geerlings_demonstrating_2013} would provide the information about the initial state populations.  However, this process involves extra complexity for a two-fluxonium system whose $\ket{1}$-$\ket{2}$ transitions have frequency and readout visibility strongly dependent on the state of the other qubit.  In order to conveniently measure and track our qubit initialization fidelity over the long course of our experiment without requiring extra control resources, 
we make use of the entangling gate available to us, the $\text{CX}_{\pi}$ gate, to calibrate qubit populations.  
Intuitively, if the conditional Rabi drive generates the same contrast as the unconditional Rabi drive, it proves that the control qubit is in a pure $\ket{1}$ state.  A quantitative procedure can be developed around this principle to calibrate any deviation of the control qubit from the pure state.  

We stress that the lack of knowledge about the initial state purity does not represent an impediment to benchmarking the two-qubit gate: The fact that the two-qubit gate fidelity is obtained via the decay rate of the RB measurements allows extraction of the the $\text{CX}_{\pi}$ gate fidelity as long as we can measure the relative qubit populations renormalized to the initial states. 
Our joint readout scheme, to be elaborated in Supplementary Section 4, provides such measurements to certify the $\text{CX}_{\pi}$ gate fidelity regardless of the initialization fidelity.  
Nevertheless, for definiteness we report actual qubit populations as calibrated in this Section throughout the manuscript.  

The initial state populations Qubit $A$ (control) and Qubit $B$ (target) are calculated with two complementary sets of measurements, each yielding the population information for one of the two qubits. The way to design such a scheme would not be unique:  As long as we have a number of independent measurements sufficient 
to set up a system of equations with robust solutions, any of them would be applicable. 
We describe our procedure under three approximations and in the end we will discuss the caveats of these approximations. These assumptions can be stated as the following:  
1) The whole population lies within the computational space ($p_{00} + p_{01} + p_{10} +p_{11} = 1$); 2) Initial residual populations for the two qubits are independent from each other; 3) The CX gate is perfect. 

\begin{figure}[tbp]
    \centering
    \includegraphics[width=8.8cm]{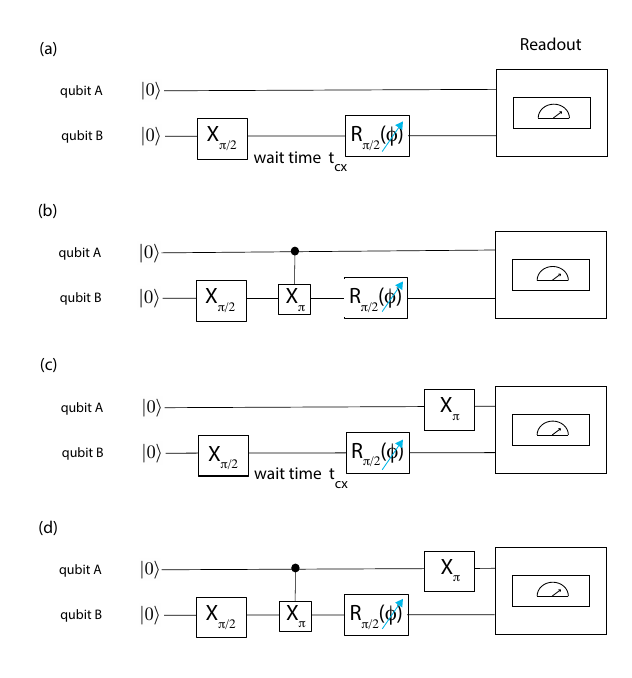}
    \caption{Pulse sequences for the set of measurements to determine the residual population of qubit A. All four measurements result in oscillations as we sweep the phase of the $\pi/2$ pulse for Qubit $B$.}
    \label{suppfig:5}
\end{figure}

\subsection{Measurement protocol for individual qubit populations}


To measure Qubit $A$'s population, we apply four Ramsey-like measurements on Qubit $B$ (Supplementary Fig.~\ref{suppfig:5}) to obtain a set of 4 equations.  In the presence of an entangling gate, the conditional flips result in different oscillation amplitudes for Qubit $B$, depending on the residual population of Qubit $A$. 
In the ideal case of no residual population for Qubit $A$ and $B$, these 4 measurements yield oscillations with the following contrasts in readout voltages:
\begin{align*}
    (a) &\xrightarrow[]{} M_{01} - M_{00}\\
    (b) &\xrightarrow[]{} M_{01} - M_{00}\\
    (c) &\xrightarrow[]{} M_{11} - M_{10}\\
    (d) &\xrightarrow[]{} M_{11} - M_{10}
\end{align*}
where $M_{ij}$ is the demodulated complex readout voltage for the pure $\ket{ij}$ state.  For this ideal case Qubit $A$ has all the population in the ground state so the contrasts $(a)-(b)$ and $(c)-(d)$ are same: There will be no difference between the unconditional and conditional oscillations since $\textsc{CX}_\pi$ has no effect on Qubit $B$ (target) with Qubit $A$ (control) being in ground. \\

Now we can 
take into account that Qubit $A$ has some non-zero population $e$ in its excited state:
\begin{align*}
    (a) &\xrightarrow[]{} (1-e)(M_{01}-M_{00}) +  e(M_{11}-M_{10}) \\
    (b) &\xrightarrow[]{} (1-e)(M_{01}-M_{00}) - e(M_{11}-M_{10})\\
    (c) &\xrightarrow[]{} (1-e)(M_{11}-M_{10}) +e (M_{01}-M_{00})\\
    (d) &\xrightarrow[]{} (1-e)(M_{11}-M_{10})-e (M_{01}-M_{00})
\end{align*}

The voltages $M_{ij}$ can be eliminated by a correct manipulation of these 4 equations:
\begin{equation}
e = \frac{(c)-(d)}{(a)+(b)+(c)-(d)}  \;\;\;\text{or}\;\;\; 
e = \frac{(a)-(b)}{(a)-(b)+(c)+(d)}
\end{equation}
to give the residual population for the control qubit. 

Qubit B's initial excited state population $\epsilon$ should also be taken into account to complete this approach.  However, it merely brings an overall $(1-2\epsilon)$ factor to all four equations that gets cancelled out during the elimination (S1). 


The population $\epsilon$ of Qubit B can be measured in a completely analogous manner with these four sets of sequences if we have a $\text{CX}_{\pi}$ gate in the reversed direction, with B as control and A as target.  While this CX$_{B\rightarrow A}$ gate is not native to our system, it can be constructed by the native CX$_{A\rightarrow B}$ gate and single-qubit Hadamard gates. Given that $\textsc{CX}_{A\rightarrow B}$ and $\textsc{CNOT}_{A\rightarrow B}$ only vary by an S-gate on Qubit $A$ (which is accounted for with a virtual-Z in the software), we apply the decomposition (CNOT$_{B\rightarrow A})=\textsc{H}_A \textsc{H}_B( \textsc{CNOT}_{A\rightarrow B}) \textsc{H}_A \textsc{H}_B$ for reverting the direction of our gate.

\subsection{Initial state populations}
Using the procedure above, we have measured the residual excited-state populations of both qubits after applying the reset protocol detailed in Supplementary Section 3.  Over the course of multiple weeks we obtain the average residual population $e$ of Qubit $A$ to be $1.0\pm0.3\%$, 
and the average residual population $\epsilon$ of Qubit $B$ to be $5.0\pm1.3\%$. 
If we assume the initial qubit states are uncorrelated, 
we can report the average initial state population after the reset protocol to be $\vec{P}_i=[P_{00}, P_{01}, P_{10}, P_{11}] \approx [0.94, 0.05, 0.01, 0.00]$. It is worth considering our assumptions and possible caveats to this initial state population measurements: 

First, since the $\text{CX}_{\pi}$ gate is not perfect, the measurement outcome of $(b)$ and $(d)$ (from Supplementary Fig. \ref{suppfig:5}) may be subject to a relative error bounded by the infidelity of the CX gate, $1-\mathcal{F}$. From Eq.~(S1), recall that $(a)\approx(b)$ and $(c)\approx(d)$ (and $(a)$ and $(c)$ are similar in magnitude), and so the possible absolute error of the calculated population $e$ and $\epsilon$ is bounded by $\frac{1}{2}(1-\mathcal{F})$, or $0.005$ for a 99\% gate fidelity.

Second, it is possible that the initial qubit states are correlated after our side-band cooling mechanism.  However, the most correlated scenario of our initial state given the data would be $\vec{P}_i \approx [0.95, 0.04, 0.00, 0.01]$, which corresponds to a change comparable to the measurement noise level. Essentially, when one of the qubit state is proven to be sufficiently pure, it strongly limits the impact of possible correlations.

Third, the initial states may have residual population in non-computational states, e.g.~in $\ket{2}$.  This population is expected to be very small considering that we managed to deplete the $\ket{1}$ state efficiently in our initialization and that the $\ket{1}$-$\ket{2}$ transition has high frequency and short relaxation time.  In addition, if Qubit $A$ is in $\ket{2}$, the Rabi oscillation of $B$ under CX drive would register distinct signature (i.e.~oscillations of a different period) that is entirely absent in the experiment (to sub-1\% level).  We cannot place as stringent of a bound to the possibility of Qubit $B$ starting in its $\ket{2}$ state, but this small population would not affect the experiment in any way expect for the overall scale of readout voltages.  For all purposes, we may simply consider $\vec{P}$ as the population vector normalized within the computation subspace.

\section{State preparation with multi-photon cooling}
\begin{figure*}[htp]
    \centering
    \includegraphics[width=13cm]{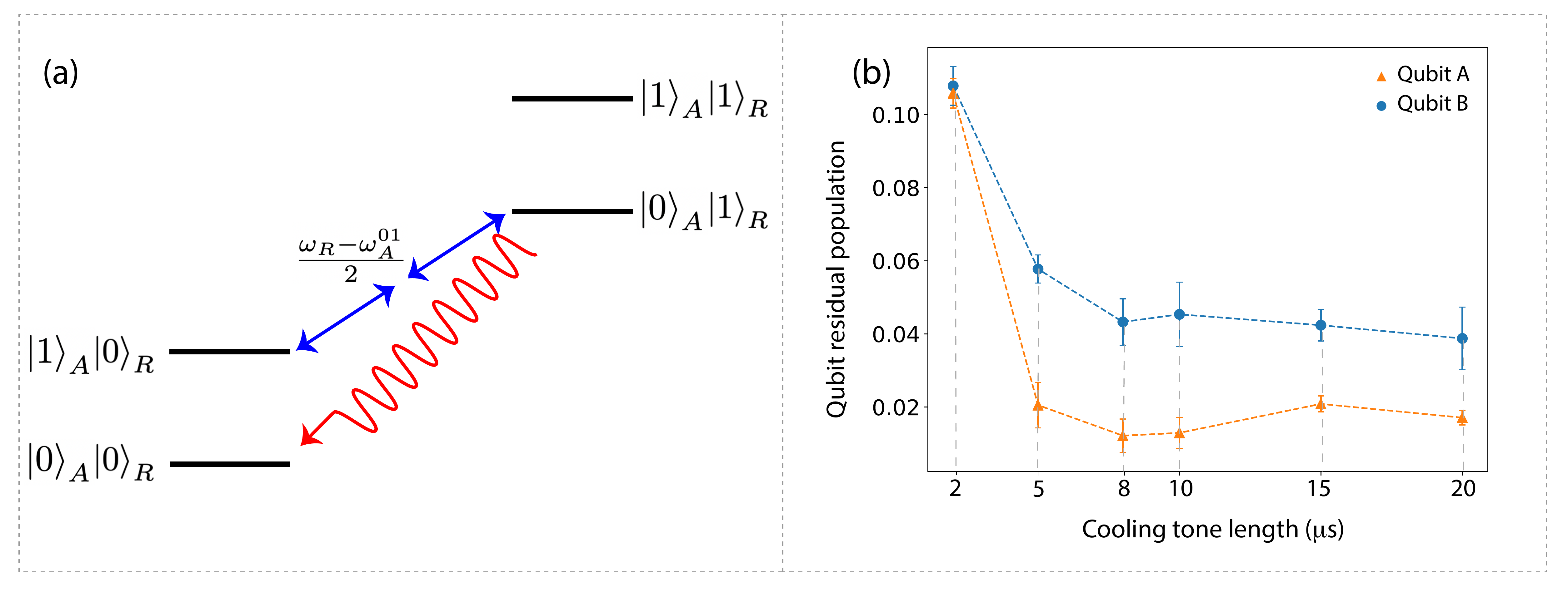}
    \caption{(a)Multi-photon cooling scheme used for initial state preparation (b) Effect of cooling tones with various lengths on qubit residual populations for the tone power and frequency specified in Supplementary Section 4 text. Each point corresponds to the average of 10 measurements (with 5000 averages each) done with same cooling configuration and the error bars are the standard error for the distribution of 10 measurements.}
    \label{suppfig:6}
\end{figure*}

Our initial state preparation protocol aims for the reset of Qubit $A$: Assuming a subsystem composed of $A$ and the cavity $R$, we drive a two-photon sideband transition from the state $\ket{1}_{A}\ket{0}_{R}$ to $\ket{0}_{A}\ket{1}_{R}$ 
which then quickly decays into $\ket{0}_{A}\ket{0}_{R}$ (Supplementary ~Fig.~\ref{suppfig:6}(a)). Initially without any reset protocol, Qubit $A$ is close to a 60\%-40\% mixture state.

To enact this multi-photon reset protocol, we apply a 15 $\mu$s-long square pulse at a frequency of 3.524 GHz, which is equal to half the difference between the cavity and Qubit $A$ frequencies. 

To calibrate the optimal cooling tone parameters, we sweep over the frequency of the cooling tone for various powers, comparing the on/off contrast of Qubit $A$. Using the qubit population measurement presented in Supplementary Section 2, we can observe the effect of the cooling tone with respect to different cooling tone lengths, reducing the residual excited state population of $A$ to $\sim$1\% and saturating after about 10 $\mu$s.  For reasons yet to be understood, this reset protocol simultaneously also resets Qubit $B$ to $\sim$5\% level.  We briefly attempted further cooling of Qubit $B$ with a $\ket{1}_B\ket{0}_R\leftrightarrow\ket{0}_B\ket{1}_R$ two-photon drive, though it did not bring additional improvement to the initialization. 

\section{Joint readout with pre-pulses}

We implement a joint readout method that makes use of a number of averaged measurements to extract the population distribution of the two qubit system~\cite{filipp_two-qubit_2009, chow_detecting_2010}.  Despite increasing the time and number of measurements, this simple scheme (with no additional experimental complexity to the readout setup) makes it possible to implement two-qubit experiments with no single-shot readout. This used to be a common practice before Josephson parametric amplifiers became widely available.  

At the end of any measurement, the readout gives us a complex demodulated voltage of the cavity transmission that involves contributions from all states weighted with their corresponding populations:
$$ 
V= p_{00}M_{00} + p_{01}M_{01} + p_{10}M_{10} + p_{11}M_{11}$$
where $\vec{p}= \text{diag}[\rho] = [p_{00}, p_{01}, p_{10}, p_{11}]^{T}$ is the population distribution for any state of the two-qubit system. This population 4-vector is particularly essential for benchmarking purposes where we need to know how much of the final state is in $\ket{00}$. 
This single complex voltage value by itself is not sufficient to pin down the population vector $\vec{p}$: In order to obtain $\vec{p}$ at any moment, we first need to know about the 4-vector $\vec{M}$ where the components are complex demodulated voltages of the cavity transmission for the computational states $[M_{00}, M_{01}, M_{10}, M_{11}]$.  

We also need to have a sufficient number of independent measurements that will let us solve for the unknowns $p_{00}, p_{01}, p_{10}, p_{11}$. For that, any prepared state will be measured with four different readout configurations, varied by pre-readout single qubit rotations:\\

1) no rotation (II) 

2) $\pi$-pulse on Qubit $B$ (IX)

3) $\pi$-pulse on Qubit $A$ (XI)

4) $\pi$-pulse on both qubits (XX)\\

With this scheme, each measurement will eventually yield a set of 4 measured voltages: $\vec{V}=[V_{II}, V_{IX}, V_{XI}, V_{XX}]^T$. 
The correct mapping relates the four populations to the four measured voltages:
$$\begin{pmatrix}
V_{II}\\
V_{IX}\\
V_{XI}\\
V_{XX}\\
\end{pmatrix} = 
\begin{pmatrix}
M_{00} & M_{01} & M_{10} & M_{11}\\
M_{01} & M_{00} & M_{11} & M_{10}\\
M_{10} & M_{11} & M_{00} & M_{01}\\
M_{11} & M_{10} & M_{01} & M_{00}\\
\end{pmatrix}
\begin{pmatrix}
p_{00}\\
p_{01}\\
p_{10}\\
p_{11}\\
\end{pmatrix}$$
The matrix M is nothing but a permutation of the components of $\vec{M}$. The vector $\vec{M}$ will allow us to reconstruct the inverse of the mapping relation that will give any state's population distribution, given the measured voltages:
\begin{equation}
\vec{p} = M^{-1} \vec{V}  
\end{equation}

In the case of perfect state preparation, obtaining components of the $\vec{M}$ would be straightforward. Although it is not a realistic assumption, it is intuitive to look at the case with $\vec{P_{i}}=[1,0,0,0]^{T}$ where $\vec{P_{i}}$ is the 4-vector for the initial states of the system:
$$\begin{pmatrix}
V_{II}\\
V_{IX}\\
V_{XI}\\
V_{XX}\\
\end{pmatrix}  = 
\begin{pmatrix}
M_{00} & M_{01} & M_{10} & M_{11}\\
M_{01} & M_{00} & M_{11} & M_{10}\\
M_{10} & M_{11} & M_{00} & M_{01}\\
M_{11} & M_{10} & M_{01} & M_{00}\\
\end{pmatrix}
\begin{pmatrix}
1\\
0\\
0\\
0\\
\end{pmatrix}=
\begin{pmatrix}
M_{00}\\
M_{01}\\
M_{10}\\
M_{11}\\
\end{pmatrix}$$

For a more realistic case, we know that the initial population vector should have non-zero populations for the remaining computational states as well. With the initial state populations $\vec{P_{i}}$ (obtained using the method from Supplementary Section 2), the mapping can be written in the following way:
$$\begin{pmatrix}
M_{00}\\
M_{01}\\
M_{10}\\
M_{11}\\
\end{pmatrix} = 
\begin{pmatrix}
P_{00} & P_{01} & P_{10} & P_{11}\\
P_{01} & P_{00} & P_{11} & P_{10}\\
P_{10} & P_{11} & P_{00} & P_{01}\\
P_{11} & P_{10} & P_{01} & P_{00}\\
\end{pmatrix}^{-1}
\begin{pmatrix}
V_{II}\\
V_{IX}\\
V_{XI}\\
V_{XX}\\
\end{pmatrix}  $$
which gives us the required complex demodulated voltages. With $\vec{M}$'s components at hand, we can now construct the full mapping from equation (S2) that gives us the population distribution for the computational states at any point of a measurement.

\section{Calibration and the benchmarking of Single Qubit Gates}

\begin{figure}[htp]
    \centering
    \includegraphics[width=8cm]{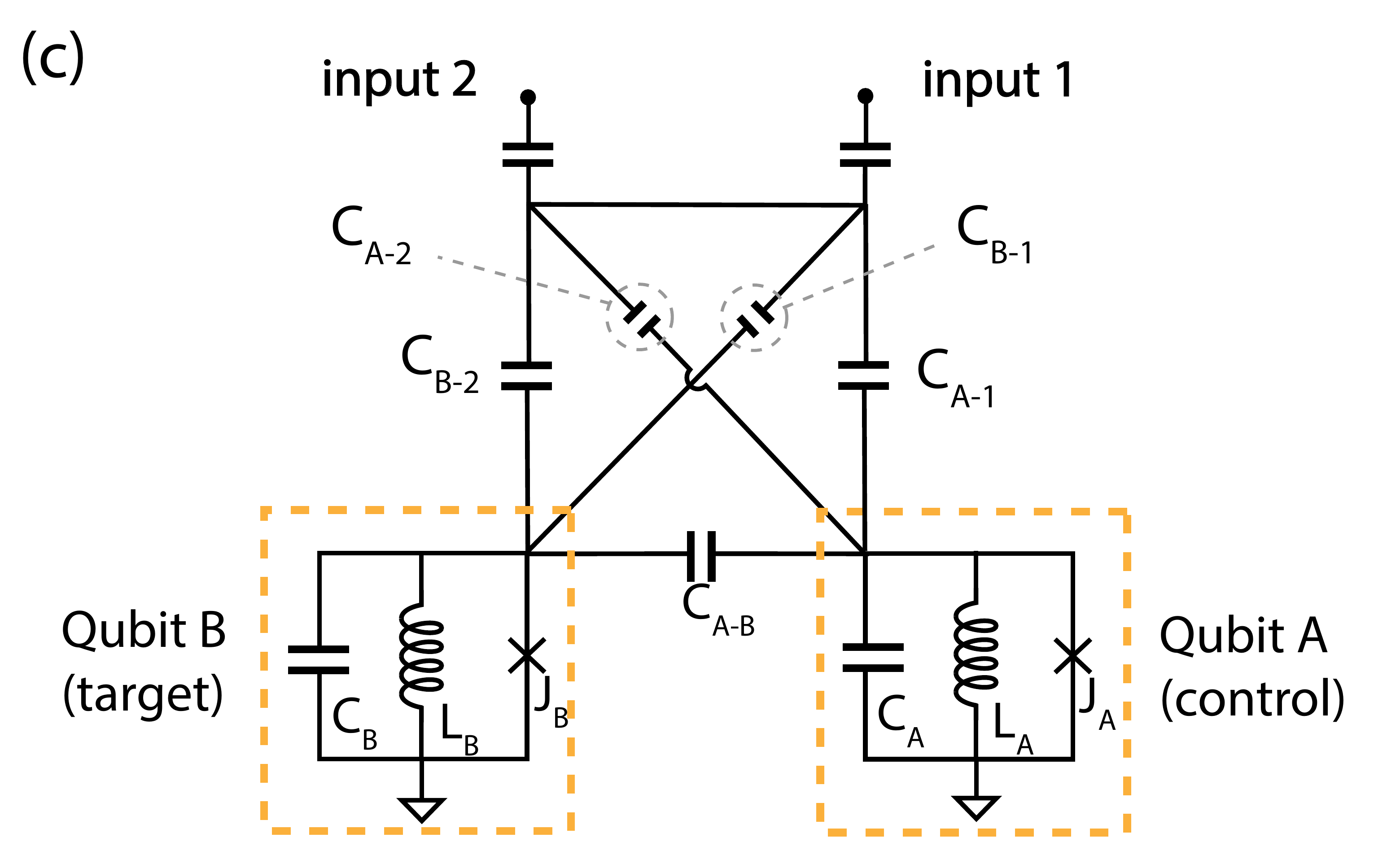}
    \caption{Effective circuit diagram of the capacitively coupled two-fluxonium system with asymmetric coupling to two input ports. The cavity mode is omitted for simplicity. }
    \label{suppfig:12}
\end{figure}

The fact that we are working in a 3D geometry with no local drives and with a system that has a strong CR component brings up the challenge of addressing the qubits individually: Given that both input ports are coupled to both qubits (see Supplementary Fig.~\ref{suppfig:12}), any qubit $B$ drive from one of the individual ports (at frequency $\omega_{B}$) will also be seen by Qubit $A$, leading to conditional oscillations of $B$ (Supplementary Fig.~\ref{suppfig:7}(a) and (b)). In order to realize high-fidelity single qubit gates, we need to be able to address each qubit independently from the state of the other one. As mentioned in the main text, the frequency of the Qubit $A$ oscillations driven from input 1 does not have a significant dependence on the Qubit $B$ state and can be considered as a ``clean" single qubit drive by itself. On the other hand, in order to realize a ``clean" single qubit gate for B, the two individual drives need to be combined with a scheme similar to the CR gate, only this time to fulfill the requirement: 
$\mel{00}{\hat H_{drive}}{01} = \mel{10}{\hat H_{drive}}{11}$
with the drive Hamiltonian from Eq.~(5) of the main text. With a similar procedure, one can find the right relative amplitude and phase between the two drives to satisfy this condition. This vector compensation scheme is illustrated in more detail in Supplementary Fig.~\ref{suppfig:8}. We can consider the drive from each port as having two components: a direct drive, which is independent of the state of Qubit $A$, and a CR drive, which is turned on whenever Qubit $A$ is in $\ket{1}$. We tune the complex ratio $\eta'$ of the drives on each port such that the CR components are of equal amplitude and opposite phase, thereby leaving only the direct drives. With the proper ratio, Qubit $B$'s Rabi frequency no longer depends on the state of Qubit $A$ (Supplementary Fig.~\ref{suppfig:7}(c)).

\begin{figure}[htp]
    \centering
    \includegraphics[width=8.7cm]{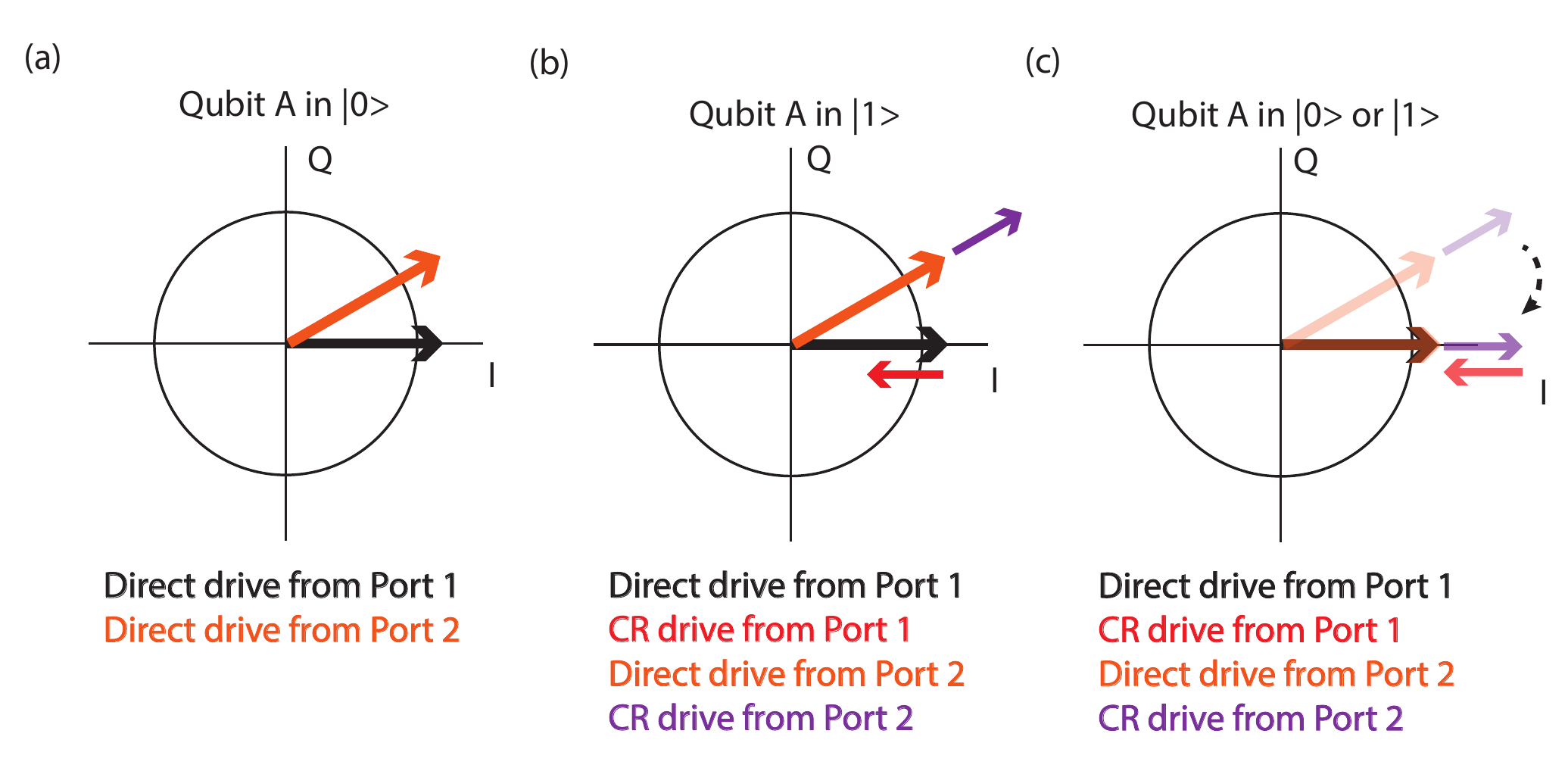}
    \caption{(a) Effective drive on qubit B, shown in the IQ plane, when qubit A is in $\ket{0}$. The qubit is driven with a different amplitude and phase depending on which driving port is used. (b) Effective drive on qubit B, shown in the IQ plane, when qubit A is in $\ket{1}$. The qubit is driven with a different effective amplitude and phase due to the CR component of the drive now being turned on.  (c) Relative amplitude and phase are applied to the two driving ports to exactly cancel the  CR drives and achieve unconditional single qubit rotations.}
    \label{suppfig:8}
\end{figure}

\begin{figure*}[htp]
    \centering
    \includegraphics[width=19cm]{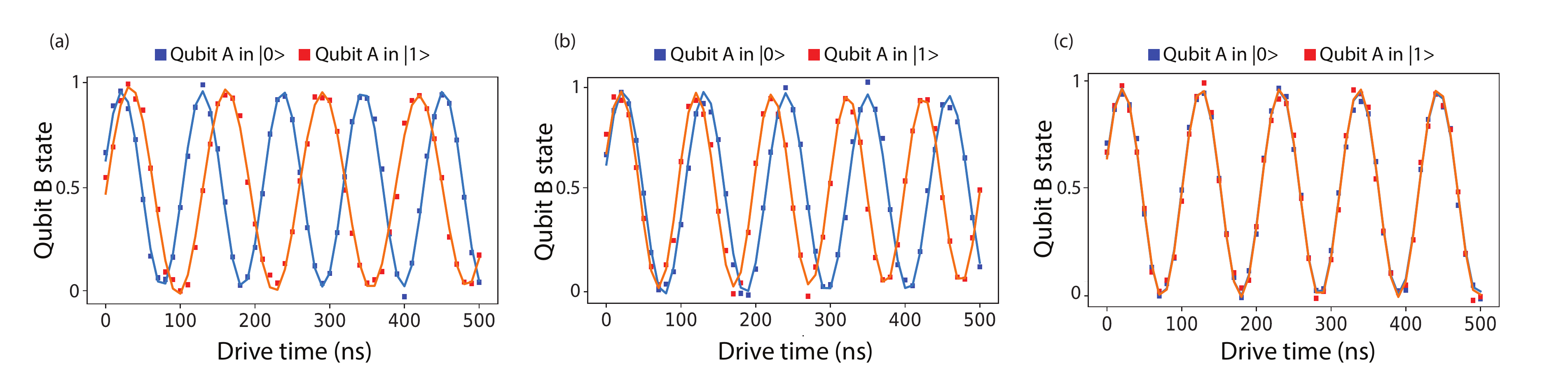}
    \caption{(a) Rabi oscillations of Qubit B driven from Port 1 (b) Rabi oscillations of Qubit B driven from Port 2 (c)
    Rabi oscillations of Qubit B after applying the complex ratio $\eta'$ between the two driving ports.}
    \label{suppfig:7}
\end{figure*}

We benchmark our single qubit gates using simultaneous single qubit randomized benchmarking: We generate random sequences of N pairs of Cliffords, one for each qubit, with N ranging from 1 to 200, and apply them to each qubit simultaneously in time. At the conclusion of each sequence, we measure two different quantities: the fidelity of the $\ket{0}$ state of qubit A and the fidelity of the $\ket{0}$ state of qubit B. This measurement was repeated 456 times over the course of several weeks, with different random sequences of N Clifford pairs being generated each time. The results were averaged together, separately for each qubit, and each fit to a single exponential. The resulting traces are shown in Supplementary Figure 9. From these traces, we can extract the average error per Clifford on each individual qubit. Taking into account the construction of our single qubit Clifford gate set, which utilizes virtual Z gates and contains 1.167 physical single qubit gates per Clifford, we report the following average fidelities per physical gate:  $\mathcal{F}_{A} = 0.99835(4)$ and $\mathcal{F}_{B} = 0.99734(6)$.

\begin{figure}[htp]
    \centering
    \includegraphics[width=8.5cm]{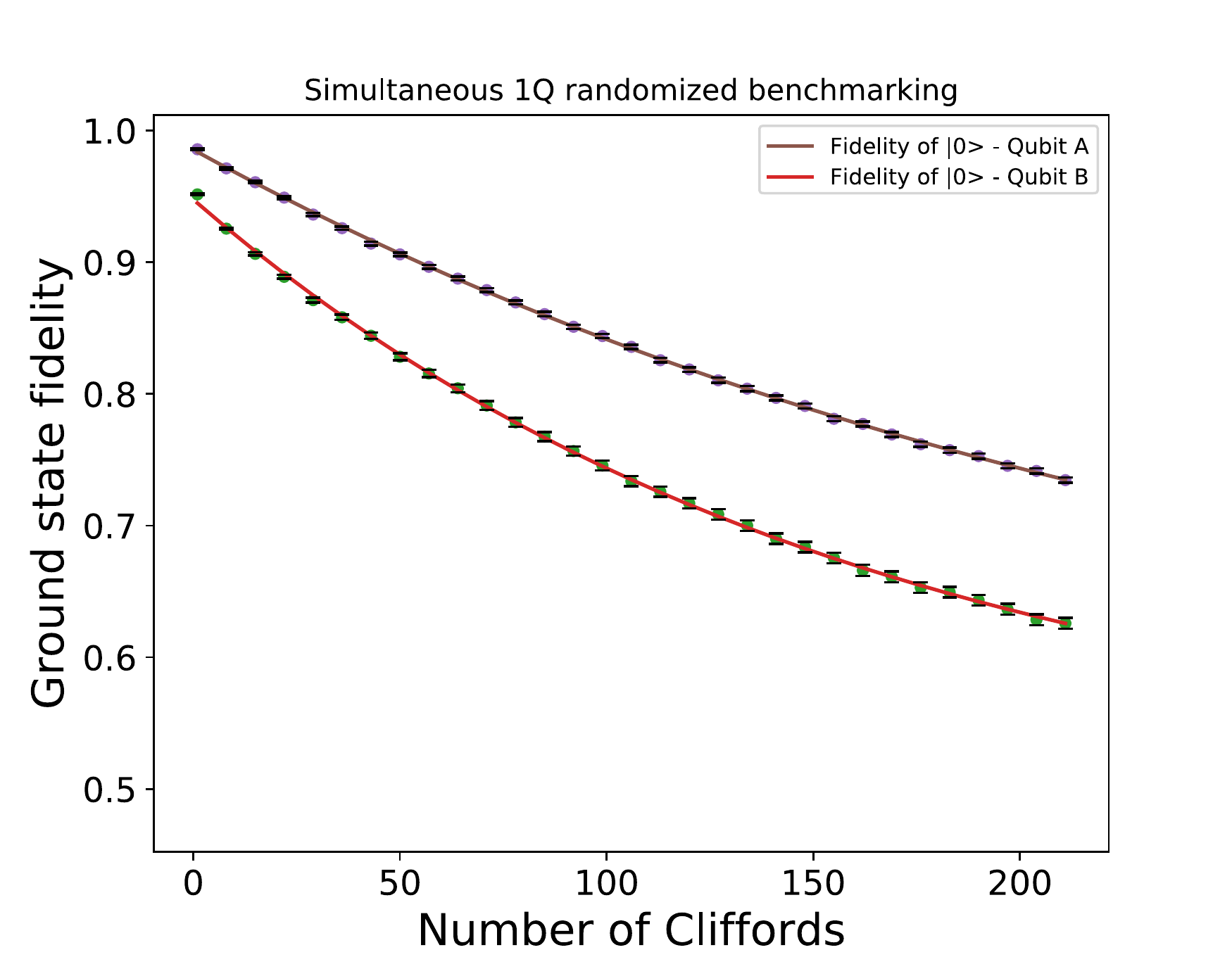}
    \caption{Simultaneous single qubit randomized benchmarking. The two traces shown are measured fidelities to the $\ket{0}$ state of qubit A and the $\ket{0}$ state of qubit B at the conclusion of N random pairs of single-qubit Clifford gates. Each trace is the averaged result of 456 randomized benchmarking experiments performed over the course of several weeks. The errorbars shown are the standard errors taken across these 456 experiments.}
\end{figure}

\section{Calibration and benchmarking of the CR gate}

The preliminary $\textsc{CX}_\pi$ gate tune up involves the calibration of the relative phase, relative amplitude and the overall amplitude, as elaborated in the main text. But we are in need of a more sophisticated calibration procedure in order to realize a high-fidelity CX gate:  Assuming that we already have a gate with the above three parameters calibrated reasonably well, we can now proceed with the next level of calibration. We should also note down that we do not have any particular gate parameter optimization procedure and these well designed tune-ups are what we rely on for realizing high-fidelity gates.

From the observed dynamics of our CR gate, we know that the process has a block-diagonal structure. The remaining degrees of freedom that would not be distinguishable for the drives from Fig. 3 of main text are the phases for the two qubits. Taking them into account, we can write our experimental CR matrix to have the following form:
\begin{equation}
 \textsc{CR}_{exp} = \begin{pmatrix}
  1 & 0 & 0 & 0 \\
  0 & e^{i\theta_{B}} & 0 & 0 \\
  0 & 0 & 0 & -ie^{i\theta_{A}} \\
  0 & 0 & -ie^{i\theta_{A}} & 0 
 \end{pmatrix}\
\end{equation}
 where $\theta_{A}$ and $\theta_{B}$ are phases picked by Qubit $A$ and $B$ during our cross-resonance gate operation. Brought in a superposition state, the phases $\theta_{A}$ ($\theta_{B}$) would be what we would observe to be the additional phase accumulated on the Qubits $A$ ($B$) under the CR drive. It is possible to measure these phases using the same pulse sequences from Supplementary Fig.~\ref{suppfig:5}: The phase difference between the oscillations resulting from the measurements of (a) and (b) on qubit A (qubit B) will give us $\theta_{A}$ ($\theta_{B})$. 
 
\subsection{Integrating single-qubit Z rotations}

Given that our experimental cross-resonance gate realization brings along extra phase accumulations on both Qubit $A$ and $B$, we first need to show the equivalence of our experimental gate to a $\textsc{CX}_\pi$ gate.

 The relevant Z-rotation operations that Qubit $A$ and $B$ undergo are expressed as:
 \begin{align*}
  \textsc{Z}^{A}(\theta_{A}) = \begin{pmatrix}
  1 & 0 & 0 & 0 \\
  0 & 1 & 0 & 0 \\
  0 & 0 & 0 & e^{i\theta_{A}} \\
  0 & 0 & e^{i\theta_{A}} & 0 
 \end{pmatrix}\ \\
\textsc{Z}^{B}(\theta_{B}) = \begin{pmatrix}
  1 & 0 & 0 & 0 \\
  0 & e^{i\theta_{B}} & 0 & 0 \\
  0 & 0 & 1 & 0 \\
  0 & 0 &0 & e^{i\theta_{B}} 
 \end{pmatrix}\
\end{align*}
It can be shown that our experimental CR gate is equivalent to a CX gate up to a number of Z-rotations:
\begin{equation}
\left[\text{CX}_{\pi}\right] = \left[\text{Z}^{B}({-\frac{\theta_{B}}{2}})\right]\left[\text{CR}_{exp}\right]\left[ \text{Z}^{B}(-\frac{\theta_{B}}{2})\right]\left[\text{Z}^{A}(\frac{\theta_{B}}{2}-\theta_{A})\right]
\label{cxequivalence}
\end{equation}
Since these Z-rotations can simply be applied via virtual Z's~\cite{mckay_efficient_2017} in the software, there is a complete equivalence between our experimental cross-resonance gate and a $\text{CX}_{\pi}$ gate as long the correct virtual Z operations are applied.  We can also trivially change $\theta_A$ by $\pi/2$ to convert the gate to a \textsc{CNOT}.  For the rest of the section and for all our benchmarking purposes, our gate is composed of the form as in Eq.~(\ref{cxequivalence}). 

\subsection{Fine tuning the gate for high-fidelity}
\begin{figure*}[tbp]
    \centering
    \includegraphics[width=15cm]{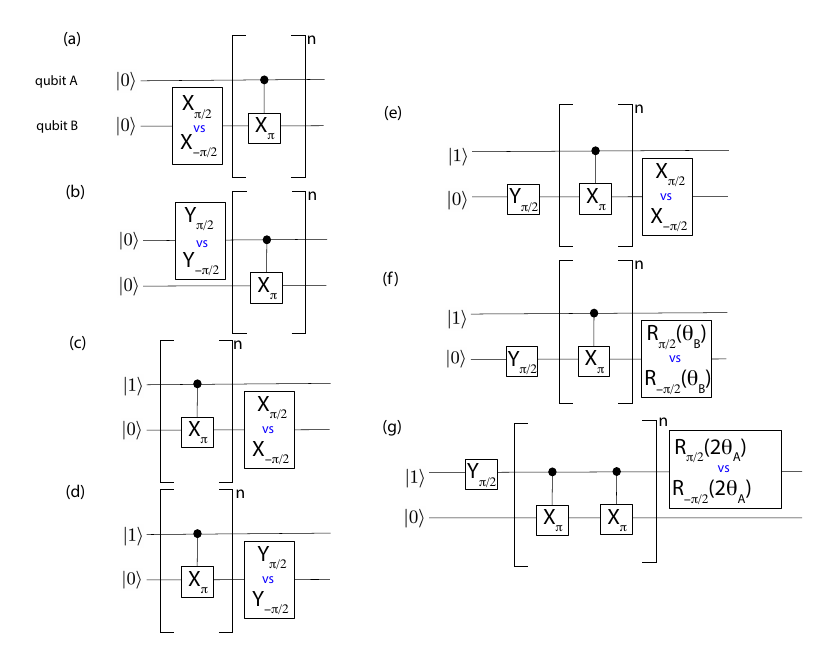}
    \caption{CR tune up sequences: These 7 experiments check the following criteria: 1) No drive on qubit B when qubit A is off: The calibration experiment (a) makes sure that qubit B does not rotate with respect to X-axis when qubit A is in state 0. Experiment (b) does a similar check for the Y-axis. 2) Resonant-$\pi$ condition when qubit A is on: Experiment (c) checks whether qubit B over-rotates or not, given qubit A is in state 1. Experiment (d) does the complementary check by making sure that qubit B is driven along the big circle with qubit A being on. 3) Alignment check: Experiment (e) checks if the CR rotation aligns with the X-axis of qubit B. 4) Correction of single qubit frames: Experiments (f) and (g) make sure that the extra phases on qubit A and qubit B are compensated in the correct way.}
    \label{suppfig:9}
\end{figure*}

The fine tune-up procedure consists of the calibration of the following parameters:
\begin{enumerate}
    \item Relative amplitude of the two drives
    \item Relative phase of the two drives 
    \item Overall amplitude of the drive
    \item Detuning of the CR drive with respect to the qubit B drive (CR detuning: $\Delta\omega_{\text{CR}}$)
    \item Misalignment of the CR drive with respect to the qubit B drive (CR angle: $\Delta\phi_{\text{CR}}$)
    \item Phase picked up by Qubit $B$ during the CR drive \item Phase picked up by Qubit $A$ during the CR drive
\end{enumerate}

Out of these parameters, the requirement for CR detuning (4) and CR angle (5) has been determined heuristically: The optimal values for these two parameters are small but non-zero ($\Delta\omega_{CR} \sim$ order of 100 kHz, $\Delta\phi_{\text{CR}} \sim$ order of $10^{-2}$ radians) and we suspect that their presence is related to the remaining underlying reflection issue. Our procedure is similar to the calibration procedure from the work \cite{wei_quantum_2021} and has been tailored to fine tune the seven degrees of freedom of our $\text{CX}_\pi$ gate. The detailed procedure and the pulse sequences can be seen from Supplementary Fig.~\ref{suppfig:9}: Sweeping over the calibration parameter, the two measurements that differ from each other by an inversion of one of their pulses result in a crossing point that gives the correct value for the each of these seven parameters. In order to increase the sensitivity of these calibrations, we can increase the number of repetitions of the CR gate in the sequence (an even number of repetitions is required for the $\theta_{A}$ calibration.)

This calibration procedure is run at regular intervals during gate benchmarking and is completed with a visual syndrome check that includes all the parameter checks simultaneously. 

\subsection{Quantum process tomography with joint readout}
The quantum process matrix of the CX$_\pi$ gate is estimated over a set of 1044 measurements with joint state readout.
We first apply the two-qubit gate on 36 initial states prepared by applying all combinations of simultaneous target and control single qubit gate: $I$, $X_{\pi/2}$, $X_{-\pi/2}$, $X_{\pi}$, $Y_{\pi/2}$, and $Y_{-\pi/2}$.
State tomography is then applied on the 36 new states following \cite{chow_universal_2012}, with 29 pre-read-out pulses and measurements from which the final density matrix is found via least-square optimization.
The CX$_\pi$ process matrix is estimated via a simultaneous least-square optimization over the 36 final density matrices.

\subsection{CX$_{\pi}$ Randomized Benchmarking}
   
To measure the fidelity of our $\text{CX}_{\pi}$ gate, we perform interleaved two-qubit randomized benchmarking. To construct our two-qubit Clifford gate set, we follow the procedures in \cite{corcoles_process_2013} and \cite{barends_superconducting_2014}, modifying it to create the two-qubit Clifford group using our $\text{CX}_{\pi}$ gate as the generator. The gate decomposition can be seen in Supplementary Fig.~\ref{suppfig:10}. Each two-qubit Clifford contains on average 6.483 physical single qubit gates and 1.5 CX gates.

We extract fidelity by measuring the sequence fidelity at at least 12 different N's with N typically going from 1 to 100. To measure sequence fidelity at each N, we generate at least 40 different random Clifford sequences, and for each random sequence we average 5000 times to reduce noise. The average sequence fidelity is fit according to the model $F = Ap^{m}+B$ (with no constraints) to obtain the decay constants and the error per clifford is $\text{EPC} = \frac{3}{4}(1-p)$. The error bars on Fig.3(a) are standard errors $ \left(\text{stdev}/\sqrt{N_{t}}\right)$ where ``stdev" is the standard deviation from the mean and $N_{t}$ is the number of runs. The uncertainties for the EPC's and fidelity numbers are calculated from the fit errors. The three different kind of EPC's ($\text{EPC}_{\text{ref}}$ for reference two qubit-RB;  $\text{EPC}_{\text{CX}}$ for interleaved RB with $\text{CX}_{\pi}$; $\text{EPC}_{\text{idle}}$ for interleaved RB with I) are calculated separately and average fidelities for these three are obtained via the following relations: 

$\text{Average 2Q Clifford fidelity} = (1 - \text{EPC}_{\text{ref}})$

$\text{Average $\text{CX}_{\pi}$ fidelity} = (1 - \text{EPC}_{\text{CX}})/(1 - \text{EPC}_{\text{ref}})$

$\text{Average Idle fidelity} = (1 - \text{EPC}_{\text{idle}})/(1 - \text{EPC}_{\text{ref}})$

\begin{figure}[tbp]
    \centering
    \includegraphics[width=8.5cm]{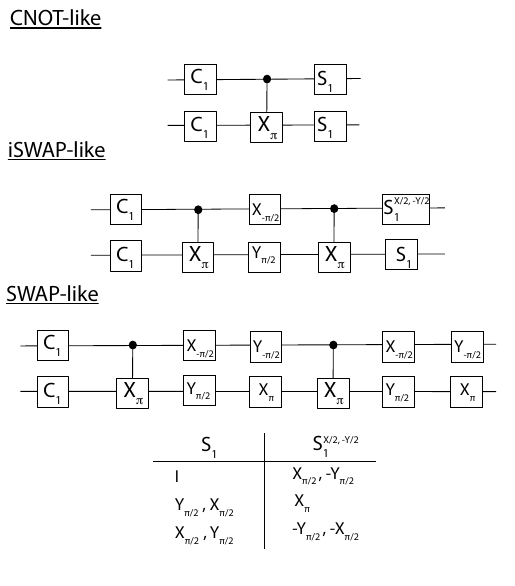}
    \caption{Two-qubit Clifford group composition with $\text{CX}_{\pi}$: $\text{C}_1$ refers to the single-qubit Clifford group, while the $\text{S}_1$ and $\text{S}_1^{X/2, -Y/2}$ are each three-element subsets of $\text{C}_1$.}
    \label{suppfig:10}
\end{figure}

\section{Experimental Setup}
\begin{figure*}[tbp]
    \centering
    \includegraphics[width=18.5cm]{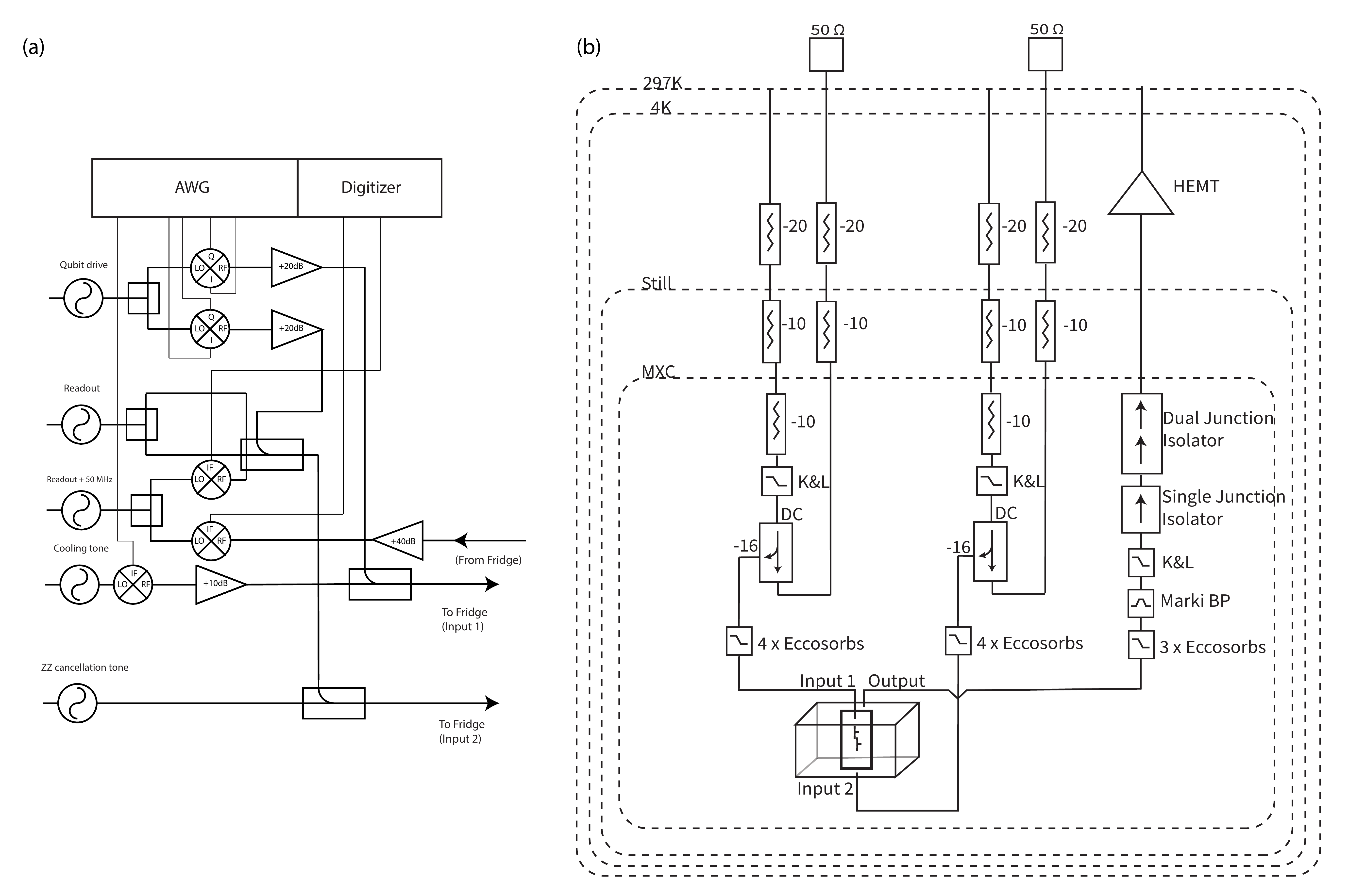}
    \caption{(a) Room temperature setup (b) Cryogenic setup}
    \label{suppfig:11}
\end{figure*}

The most fundamental microwave setup requirement for our experimental scheme is the phase locking between the two drive tones played into the two input ports. We realize this by using a single microwave generator that resources all individual qubit drives: The same local oscillator feeds the two (home-made) IQ mixers (one per control line) to maintain phase coherence during combined drives. Its frequency is set at 840 MHz based on hardware limitations and to minimize wrong sideband and qubit frequency overlaps. Each IQ mixer is made with discrete 3-port mixers and power combiners/splitters carefully chosen to cover both qubit frequencies after up-mixing (corresponding to intermediate frequencies of 164 MHz for Qubit B and -285MHz for Qubit A). The room temperature microwave setup is shown in Supplementary Fig.~\ref{suppfig:11}(a). In addition to what is shown in that figure, we also heavily filter our lines using low and band pass filters which are omitted in the diagram for simplicity.
The fridge setup is shown in Supplementary Fig.~\ref{suppfig:11}(b): The non-cryogenic directional coupler to which we attribute the impedance mismatch can be seen below the mixing chamber. All the Eccosorb filters are made in-house and have been chained up (4x for each input line and 3x for the output line) between the cavity ports and the closest microwave component. The experiment is carried out in an Oxford Triton 500 Dilution Refrigerator with a based temperature below 20 mK.

\section{Fabrication of the two-fluxonium device}

The fabrication procedure for the two-fluxonium device is identical to the one in \cite{somoroff_millisecond_2021}: The device is fabricated on a 430 $\mu$m thickness sapphire chip with dimensions $9 \times 4$ mm. Following the cleaning of the chip, the double-layer resist's first layer is prepared by spinning MMA EL-13 at 5000 RPM and baking at $180^{\circ}$C for 1 minute, followed by the second layer of the resist: 950 PMMA A3 spun at 4000 RPM for 1 minute and baked at the same temperature for half an hour. As the anti-charging layer, 11 nm of Al is deposited using a Plassys deposition system. The electron beam lithography is realized using a current of 1 nA on a 100 kV Elionix EBL system. After the electron beam lithography, the anti-charging layer is removed by keeping the chip in a 0.1 M potassium hydroxide (KOH) aqueous solution. The development is done in a 3:1 IPA:DI solution at a temperature of $6^{\circ}$C for 2 minutes. A Plassys deposition system is used for the double angle Al deposition (with a Al deposition thickness of 20 nm for the first and 40 nm for the second layer) with the same process steps detailed in \cite{somoroff_millisecond_2021}. Liftoff is realized by keeping the chip in acetone at a temperature of $60^{\circ}$C for 3 hours, followed by brief sonications in acetone and IPA (5 and 10 seconds). As the last step, the chip is dried with $N_2$.


\bibliography{References}